\documentclass[11pt]{article}
\usepackage{makeidx}
\usepackage{natbib,amsfonts,amsmath,verbatim}
\usepackage{amssymb}
\usepackage{graphicx, caption, subcaption}
\usepackage{multirow,rotating}
\usepackage{theorem}
\usepackage{fullpage}

\newtheorem{theorem}{Theorem}

\newtheorem{algorithm}{Algorithm}
\newtheorem{assumption}{Assumption}

\newtheorem{lemma}{Lemma}

\newtheorem{remark}{Remark}
\newtheorem{samplingscheme}{Sampling Scheme}

\renewcommand{\theequation}{\thesection.\arabic{equation}}

\makeatletter
\@addtoreset{equation}{section}
\makeatother
\bibpunct[,]{(}{)}{;}{a}{,}{,}
\input{tcilatex}
\begin{document}

\title{An extended space approach for particle Markov chain Monte Carlo
methods}
\author{Christopher K. Carter \and Eduardo F. Mendes \and Robert Kohn}
\maketitle

\begin{abstract}
In this paper we consider fully Bayesian inference in general state space
models. Existing particle Markov chain Monte Carlo (MCMC) algorithms use an
augmented model that takes into account all the variable sampled in a
sequential Monte Carlo algorithm. This paper describes an approach that also
uses sequential Monte Carlo to construct an approximation to the state
space, but generates extra states using MCMC runs at each time point. We
construct an augmented model for our extended space with the marginal
distribution of the sampled states matching the posterior distribution of
the state vector. We show how our method may be combined with particle
independent Metropolis-Hastings or particle Gibbs steps to obtain a
smoothing algorithm. All the Metropolis acceptance probabilities are
identical to those obtained in existing approaches, so there is no extra
cost in term of Metropolis-Hastings rejections when using our approach. The
number of MCMC iterates at each time point is chosen by the user and our
augmented model collapses back to the model in \cite{olssonryden2011} when
the number of MCMC iterations reduces. We show empirically that our approach
works well on applied examples and can outperform existing methods.
\end{abstract}

\section{Introduction\label{U:intro}}

Our article deals with statistical inference for non-Gaussian state space
models. Its main goal is to provide flexible methods that give effficient
estimates for a wide class of state space models. This work extends the
methods proposed by \cite{andrieuetal2010}, \cite{bunchgodsill2013}, \cite%
{lindstenschon2012}, \cite{lindstenetal2014} and \cite{olssonryden2011}.

MCMC methods for Bayesian inference for Gaussian state space models or
conditionally Gaussian state space models are well developed with algorithms
to generate from the joint distribution of all the state vectors and to
generate from marginal distributions with the state vectors integrated out
-- see, for example, \cite{carterkohn1994}, \cite{fruhwirth1994}, \cite%
{gerlachetal2000} and \cite{fruhwirth2006}. Bayesian inference for general
non-Gaussian state space models has proved to be a much harder problem. MCMC
approaches include single-site updating\ of the state vectors in \cite%
{carlinetal1992} and block-updating of the state vectors in \cite%
{shephardpitt1997}. These approaches apply to general models, but they can
be inefficient for some cases and can require numerical approximations over
high dimensional spaces. MCMC methods based on the particle filter have
proved to be an attactive alternative. A class of MCMC methods involving
unbiased estimation of the likelihood was introduced by \cite{beaumont2003}
and its theoretical properties are discussed in \cite{andrieuroberts2009}.

\cite{andrieuetal2010} extend these methods by constructing a joint
distribution for the output of the particle filter that has a marginal
distribution equal to the posterior distribution of the states in a state
space model. This marginal distribution involves the states determined by
tracing back the ancestors of a selected particle and is called the \textit{%
ancestral tracing} approach by \cite{andrieuetal2010}. They show that
previous approaches involving unbiased estimation of the likelihood
correspond to Metropolis-Hastings sampling schemes under their joint
distribution. The methods in \cite{andrieuetal2010} can also be viewed as a
fully Bayesian approach to the smoothing algorithm of \cite{kitagawa1996}.
The \cite{andrieuetal2010} approach also allows other possible MCMC sampling
schemes and they construct a particle Gibbs sampler which targets the same
joint distribution. \cite{lindstenetal2014} construct another particle Gibbs
sampler for this model and give empirical evidence that their sampler
improves the mixing properties of the resulting Markov chains. \cite%
{dubarrydouc2012} give a smoothing method based on single-site MCMC updating
of the generated trajectories from the ancestral tracing approach in \cite%
{andrieuetal2010}.

\cite{olssonryden2011} extend the methods in \cite{andrieuetal2010} by
contructing a joint distribution on the ouput of the particle filter
together with a series of indices corresponding to the selected states. The
sampling of indices is based on the forward filtering backward simulation
approach in \cite{godsilletal2004} and is called the \textit{backward
simulation} approach in the literature. Their joint distribution also has a
marginal distribution equal to the posterior distribution of the states in a
state space model and their Metropolis-Hastings sampling schemes have the
same acceptance probabilities as the \cite{andrieuetal2010} approach. \cite%
{lindstenschon2012} constructs a particle Gibbs algorithm for the \cite%
{olssonryden2011} model and gives empirical results showing improved
effciency over previous approaches. \cite{chopinsingh2013} gives theoretical
results showing the particle Gibbs with backward simulation in \cite%
{lindstenschon2012} has a smaller integrated autocorrelation time compared
to the \cite{andrieuetal2010} particle Gibbs sampler.

\cite{bunchgodsill2013} give a smoothing algorithm which runs the particle
filter and then uses a backwards simulation approach that involves running
an MCMC at each time point. They show that the advantage of their method is
that new values of the state vectors are generated during the backward
simulation step, whereas many other approaches are restricted to the output
of the particle filter. \cite{fearnheadetal2010} give a smoothing algorithm
based on combining particles from a forward filter and a backward
information filter, which also generates new values of the state vectors.

Our work extends the methods in \cite{olssonryden2011}, \cite%
{lindstenschon2012} \ and \cite{bunchgodsill2013} by using an augmented
model that includes the results of the particle filter, a series of indices
which correspond to starting values of an MCMC run at each time point, and
the output of the MCMC runs. We construct a joint distribution for our
augmented space which has a marginal distribution equal to the posterior
distribution of the states in a state space model and we show that our
Metropolis-Hastings sampling schemes have the same acceptance probabilities
as the approaches in \cite{andrieuetal2010} and \cite{olssonryden2011}. The
advantage of our approach is that the MCMC runs at each time point generate
new values of the state vectors, so we are not restricted to the output of
the particle filter. Our method can be used to obtain generated states from
the smoothing distrution or for Bayesian inference involving parameters. Our
method is fully Bayesian, so the output of our MCMC convergences to the
posterior distribution given suitable regularity conditions which we
discuss. We derive a particle Gibbs sampler for our augmented model.

The paper is organised as follows. Section \ref{U:genstates} describes our
state space model and sequential Monte Carlo algorithm. This section also
constructs the joint distribution we use for Bayesian inference, describes
the properties of this distubution, and gives our particle Gibbs algorithm.
Section \ref{U:Estimation} describes our MCMC sampling schemes to carry out
smoothing and Bayesian inference and discusses their convergence properties.
Section \ref{U:Examples} reports the empirical results. Proofs are given in
an Appendix.

\section{Generating the states \label{U:genstates}}

This section gives the technical results that are required for the Markov
chain Monte Carlo methods described in Section \ref{U:Estimation}. We
describe the State Space Model, the Sequential Monte Carlo algorithm to
generate the particles, and the extra Markov chain Monte Carlo steps in our
method to generate the states. We then derive the properties of the
distributions resulting from our algorithms. We also give a conditional
sequential Monte Carlo algorithm that is used for particle Gibbs steps in
Section \ref{U:Estimation}. We use the standard convention where capital
letters denote random variables and lower case letters denote\ their values.

\subsection{State Space Model\label{U:SSM}}

Consider the state space model with states denoted by $\left\{
X_{t}:t=1,\ldots ,T\right\} \subset \mathcal{X}^{T}$and observations denoted
by $\left\{ Y_{t}:t=1,\ldots ,T\right\} $. We will assume the transition and
observation distributions have positive densities denoted by%
\begin{eqnarray}
X_{1} &\sim &f_{1}\left( \cdot |\theta \right) \text{ \ \ \ and}
\label{ssm_1} \\
X_{t}|\left( X_{t-1}=x\right) &\sim &f_{t}\left( \cdot |x,\theta \right) 
\text{ \ \ }t=2,\ldots ,T  \label{ssm_2} \\
Y_{t}|\left( X_{t}=x\right) &\sim &g_{t}\left( \cdot |x,\theta \right) \text{
\ \ }t=1,\ldots ,T.  \label{ssm_3}
\end{eqnarray}%
All the densities are with respect to Lebesgue measure for continuous
variables and counting measure for discrete valued variables unless
otherwise indicated. The vector $\theta \in \Theta $ represents parameters
which are discussed in Section \ref{U:Estimation} and in the examples in
Section \ref{U:Examples}. We use the following notation for sequences, $%
\,z_{i:j}=\left( z_{i},\ldots ,z_{j}\right) $ and we denote the joint
density of $\left\{ y_{1:T},x_{1:T}\right\} $ given $\theta $ by%
\begin{equation}
p\left( y_{1:T},x_{1:T}|\theta \right) \mathrel{:=}g_{1}\left(
y_{1}|x_{1},\theta \right) f_{1}\left( x_{1}|\theta \right)
\dprod_{t=2}^{T}g_{t}\left( y_{t}|x_{t},\theta \right) f_{t}\left(
x_{t}|x_{t-1},\theta \right) .  \label{ssm_4}
\end{equation}

\subsection{Sequential Monte Carlo algorithm\label{U:SMC}}

The Sequential Monte Carlo algorithm we use for the state space model\
defined by (\ref{ssm_1})--(\ref{ssm_4}) at time $t$ constructs a sample of
particles denoted by $\left\{ X_{1:t}^{1},\ldots ,X_{1:t}^{N_{t}}\right\} $
with associated normalized weights $\mathbf{W}_{t}=\left\{ W_{t}^{1},\ldots
,W_{t}^{N_{t}}\right\} $ that approximates the distribution $p\left(
dx_{1:t}|y_{1:t},\theta \right) $ by%
\begin{equation}
\widehat{p}\left( dx_{1:t}|y_{1:t},\theta \right) \mathrel{:=}%
\dsum_{i=1}^{N_{t}}W_{t}^{i}\delta _{X_{1:t}^{i}}\left( dx_{1:t}\right) .
\label{smc_1}
\end{equation}%
In the pseudocode of the sequential Monte Carlo Algorithm \ref{SMC_algorithm}
described below we denote the unnormalized weights at time $t$ by $\mathbf{w}%
_{t}=\left\{ w_{t}^{1},\ldots ,w_{t}^{N_{t}}\right\} $ and use the notation $%
\mathcal{F}\left( \cdot |\mathbf{p}\right) $ for the discrete probability
distribution on $\left\{ 1,\ldots ,m\right\} $ of parameter $\mathbf{p}%
=\left( p_{1},\ldots ,p_{m}\right) $, with $p_{i}\geq 0$ and $p_{1}+\ldots
+p_{m}=1$, for some $m\in 
\mathbb{N}
$. Algorithm \ref{assumption:support} uses the importance densities $%
M_{1}\left( x_{1}|y_{1},\theta \right) $ and $M_{t}\left(
x_{t}|y_{t},x_{t-1},\theta \right) ,$ for $t=2,\ldots ,T$. We make
Assumption \ref{assumption:support} about these importance densities for the
results in later sections.

\begin{assumption}
\label{assumption:support} $M_{1}\left( x_{1}|y_{1},\theta \right) $ and $%
M_{t}\left( x_{t}|y_{t},x_{t-1},\theta \right) ,$ for $t=2,\ldots ,T$ are
finite strictly positive densities.
\end{assumption}

Algorithm \ref{SMC_algorithm} is based on \cite{andrieuetal2010} and we
include it for completeness and notational consistency. We use the
convention that whenever the index $i$ is used for a particular value of $t$
we mean `for all $i\in \left\{ 1,\ldots ,N_{t}\right\} $'.

\begin{algorithm}[Sequential Monte Carlo]
\label{SMC_algorithm}\ \ 

\begin{description}
\item[Step 1] For $t=1,$

\begin{description}
\item[Step 1.1] sample $X_{1}^{i}\sim M_{1}\left( \cdot |y_{1},\theta
\right) ,$

\item[Step 1.2] compute and normalize the weights%
\begin{equation}
w_{1}^{i}=\frac{p\left( y_{1}|X_{1}^{i},\theta \right) p\left(
X_{1}^{i}|\theta \right) }{M_{1}\left( X_{1}^{i}|y_{1},\theta \right) }.
\label{smc_2}
\end{equation}%
\begin{equation*}
W_{1}^{i}=\frac{w_{1}^{i}}{\dsum_{j=1}^{N_{1}}w_{1}^{j}}.
\end{equation*}
\end{description}

\item[Step 2] For $t=2,\ldots ,T,$

\begin{description}
\item[Step 2.1] sample $A_{t-1}^{i}\sim \mathcal{F}\left( \cdot |\mathbf{W}%
_{t-1}\right) ,$

\item[Step 2.2] sample $X_{t}^{i}\sim M_{t}\left( \cdot
|y_{t},X_{t-1}^{A_{t-1}^{i}},\theta \right) ,$

\item[Step 2.3] compute and normalize the weights%
\begin{equation}
w_{t}^{i}=\frac{p\left( y_{t}|X_{t}^{i},\theta \right) p\left(
X_{t}^{i}|X_{t-1}^{A_{t-1}^{i}},\theta \right) }{M_{t}\left(
X_{t}^{i}|y_{t},X_{t-1}^{A_{t-1}^{i}},\theta \right) }.  \label{smc_3}
\end{equation}%
\begin{equation*}
W_{t}^{i}=\frac{w_{t}^{i}}{\dsum_{j=1}^{N_{t}}w_{t}^{j}}.
\end{equation*}
\end{description}
\end{description}
\end{algorithm}

The variable $A_{t-1}^{i}$ in the above algorithm represents the index of
the parent at time $t-1$ of particle $X_{1:t}^{i}$. Our methods do not
require the full trajectory of the states in a particle and are more
concerned with the individual values $X_{t}^{i}$ for $t=1,\ldots ,T$ and $%
A_{t}^{i}$ for $t=1,\ldots ,T-1$. We denote the collection of states at time 
$t$ by $\overline{\mathbf{X}}_{t}=\left\{ X_{t}^{i}:i=1,\ldots
,N_{t}\right\} $ for $t=1,\ldots ,T$ and the corresponding collection of
parent indices by $\overline{\mathbf{A}}_{t}=\left\{ A_{t}^{1},\ldots
,A_{t}^{N_{t}}\right\} $ for $t=1,\ldots ,T-1$. We will also use the
notation for sequences $\overline{\mathbf{X}}_{i:j}=\left( \overline{\mathbf{%
X}}_{i},\ldots ,\overline{\mathbf{X}}_{j}\right) $ and $\overline{\mathbf{A}}%
_{i:j}=\left( \overline{\mathbf{A}}_{i},\ldots ,\overline{\mathbf{A}}%
_{j}\right) $.

\subsection{MCMC steps to generate states\label{U:MCMCGenStates}}

Algorithm \ref{MCMCGenStates_algorithm} described below takes the output of
the Sequential Monte Carlo steps described in Algorithm \ref{SMC_algorithm}
and runs a backward simulation algorithm to generate extra state values. The
state at time $T$ is generated using the approach in \cite{andrieuetal2010}
and the states at times $T-1,\ldots ,1$ are generated using an approach
related to that of \cite{bunchgodsill2013}, which, at time $t,$ involves a
Markov chain Monte Carlo run of length $C_{t}$. We denote the generated
values at time $t$ by $\widetilde{\mathbf{X}}_{t}=\left\{ \tilde{X}%
_{t}^{i}:i=1,\ldots ,C_{t}\right\} $ for $t=1,\ldots ,T-1$ and use the
sequence notation $\widetilde{\mathbf{X}}_{i:j}=\left( \widetilde{\mathbf{X}}%
_{i},\ldots ,\widetilde{\mathbf{X}}_{j}\right) .$

These Markov chain Monte Carlo runs involve the following components. For $%
t=1$, the target density for the Metropolis-Hasting step is%
\begin{equation*}
p\left( x_{1}|y_{1,}\tilde{x}_{2}^{C_{2}},\theta \right) \varpropto
g_{1}\left( y_{1}|x_{1},\theta \right) f_{2}\left( \tilde{x}%
_{2}^{C_{2}}|x_{1},\theta \right) ,
\end{equation*}%
so no approximation using the output from Algorithm \ref{SMC_algorithm} is
required. For $t=2,\ldots ,T-2$, the target densities for the
Metropolis-Hasting steps are%
\begin{equation}
\widehat{p}\left( x_{t}|\overline{\mathbf{x}}_{1:t-1},\tilde{x}%
_{t+1}^{C_{t+1}},\overline{\mathbf{a}}_{1:t-2},\theta \right) \varpropto
g_{t}\left( y_{t}|x_{t},\theta \right) f_{t+1}\left( \tilde{x}%
_{t+1}^{C_{t+1}}|x_{t},\theta \right)
\dsum_{i=1}^{N_{t-1}}w_{t-1}^{i}f_{t}\left( x_{t}|x_{t-1}^{i},\theta \right)
,  \label{MCMCGenStates_1a}
\end{equation}%
which approximates%
\begin{eqnarray*}
&&p\left( x_{t}|y_{1:t},X_{t+1}=\tilde{x}_{t+1}^{C_{t+1}},\theta \right)
\varpropto g_{t}\left( y_{t}|x_{t},\theta \right) f_{t+1}\left( \tilde{x}%
_{t+1}^{C_{t+1}}|x_{t},\theta \right) \times \\
&&\dint p\left( x_{t}|y_{1:t-1},x_{t-1},\theta \right) p\left(
x_{t-1}|y_{1:t-1},\theta \right) dx_{t-1}
\end{eqnarray*}%
based on the output from Algorithm \ref{SMC_algorithm}. Similarly, for $%
t=T-1 $ the target density for the Metropolis-Hasting steps is%
\begin{equation}
\widehat{p}\left( x_{T-1}|\overline{\mathbf{x}}_{1:T-2},\tilde{x}%
_{T}^{b_{T}},\overline{\mathbf{a}}_{1:T-3},\theta \right) \varpropto
g_{t}\left( y_{T-1}|x_{T-1},\theta \right) f_{T}\left(
x_{T}^{b_{T}}|x_{T-1},\theta \right)
\dsum_{i=1}^{N_{T-2}}w_{T-2}^{i}f_{T-1}\left( x_{T-1}|x_{T-2}^{i},\theta
\right) ,  \label{MCMCGenStates_1aa}
\end{equation}%
which approximates%
\begin{eqnarray*}
&&p\left( x_{T-1}|y_{1:T-1},X_{T}=x_{T}^{b_{T}},\theta \right) \varpropto
g_{T-1}\left( y_{T-1}|x_{T-1},\theta \right) f_{T}\left(
x_{T}^{b_{T}}|x_{T-1},\theta \right) \times \\
&&\dint p\left( x_{T-1}|y_{1:T-2},x_{T-2},\theta \right) p\left(
x_{T-2}|y_{1:T-2},\theta \right) dx_{T-2}.
\end{eqnarray*}

The following lemma follows immediately from the assumption that $%
g_{t}\left( y_{t}|x_{t},\theta \right) $ and $f_{t}\left(
x_{t}|x_{t-1},\theta \right) $ are strictly positive densities for $%
t=1,\ldots ,T$.

\begin{lemma}
\label{MCMCGenStates:lemma_1}The densities $\widehat{p}\left( x_{t}|%
\overline{\mathbf{x}}_{1:t-1},\tilde{x}_{t+1}^{C_{t+1}},\overline{\mathbf{a}}%
_{1:t-2},\theta \right) $ for $t=2,\ldots ,T-1$ and $\widehat{p}\left(
x_{T-1}|\overline{\mathbf{x}}_{1:T-2},x_{T}^{b_{T}},\overline{\mathbf{a}}%
_{1:T-3},\theta \right) $ are strictly positive.
\end{lemma}

We denote the MCMC transition kernels by%
\begin{equation}
K_{t}\left( x_{t},dx_{t}^{\prime }|\overline{\mathbf{x}}_{1:t-1},\overline{%
\mathbf{x}}_{t}^{-b_{t}},\tilde{x}_{t+1}^{C_{t+1}},\overline{\mathbf{a}}%
_{1:t-2},\mathbf{a}_{t-1}^{-b_{t}},\theta \right) ,  \label{MCMCGenStates_2a}
\end{equation}%
for $t=1,\ldots ,T-2$ and%
\begin{equation}
K_{T-1}\left( x_{T-1},dx_{T-1}^{\prime }|\overline{\mathbf{x}}_{1:T-2},%
\overline{\mathbf{x}}_{T-1}^{-b_{t}},x_{T}^{b_{T}},\overline{\mathbf{a}}%
_{1:T-3},\mathbf{a}_{T-2}^{-b_{T-1}},\theta \right) .
\label{MCMCGenStates_2aa}
\end{equation}%
The choice of Metropolis-Hastings proposal is determined by the user, but
the conditioning indicated in (\ref{MCMCGenStates_2a}) and (\ref%
{MCMCGenStates_2aa}) is sufficient for the results given in Sections \ref%
{U:Distrib} and \ref{U:PG}. We require the standard reversibility condition
of detailed balance as described in Assumption \ref%
{assumption:detailedbalance}. Sections \ref{U:Ergodicity} and \ref%
{U:Proposal} give more detail on the transition kernels.

\begin{assumption}[Detailed balance]
\label{assumption:detailedbalance}For all $\theta \in \Theta $

\begin{description}
\item[(a)] 
\begin{equation*}
p\left( dx_{1}|y_{1,}\tilde{x}_{2}^{C_{2}},\theta \right) K_{1}\left(
x_{1},dx_{1}^{\prime }|\overline{\mathbf{x}}_{1}^{-b_{1}},\tilde{x}%
_{2}^{C_{2}},\theta \right) =p\left( dx_{1}^{\prime }|y_{1,}\tilde{x}%
_{2}^{C_{2}},\theta \right) K_{1}\left( x_{1}^{\prime },dx_{1}|\overline{%
\mathbf{x}}_{1}^{-b_{1}},\tilde{x}_{2}^{C_{2}},\theta \right) ,
\end{equation*}

\item[(b)] 
\begin{eqnarray*}
&&\widehat{p}\left( dx_{t}|\overline{\mathbf{x}}_{1:t-1},\tilde{x}%
_{t+1}^{C_{t+1}},\overline{\mathbf{a}}_{1:t-2},\theta \right) K_{t}\left(
x_{t},dx_{t}^{\prime }|\overline{\mathbf{x}}_{1:t-1},\overline{\mathbf{x}}%
_{t}^{-b_{t}},\tilde{x}_{t+1}^{C_{t+1}},\overline{\mathbf{a}}_{1:t-2},%
\mathbf{a}_{t-1}^{-b_{t}},\theta \right) \\
&=&\widehat{p}\left( dx_{t}^{\prime }|\overline{\mathbf{x}}_{1:t-1},\tilde{x}%
_{t+1}^{C_{t+1}},\overline{\mathbf{a}}_{1:t-2},\theta \right) K_{t}\left(
x_{t}^{\prime },dx_{t}|\overline{\mathbf{x}}_{1:t-1},\overline{\mathbf{x}}%
_{t}^{-b_{t}},\tilde{x}_{t+1}^{C_{t+1}},\overline{\mathbf{a}}_{1:t-2},%
\mathbf{a}_{t-1}^{-b_{t}},\theta \right) ,
\end{eqnarray*}

for $t=2,\ldots ,T-2$ and

\item[(c)] 
\begin{eqnarray*}
&&\widehat{p}\left( dx_{T-1}|\overline{\mathbf{x}}_{1:T-2},x_{T}^{B_{T}},%
\overline{\mathbf{a}}_{1:T-3},\theta \right) K_{t}\left(
x_{T-1},dx_{T-1}^{\prime }|\overline{\mathbf{x}}_{1:T-2},\overline{\mathbf{x}%
}_{T-1}^{-b_{T-1}},x_{T}^{B_{T}},\overline{\mathbf{a}}_{1:T-3},\mathbf{a}%
_{T-2}^{-b_{t}},\theta \right) \\
&=&\widehat{p}\left( dx_{T-1}^{\prime }|\overline{\mathbf{x}}%
_{1:T-2},x_{T}^{B_{T}},\overline{\mathbf{a}}_{1:T-3},\theta \right)
K_{t}\left( x_{T-1}^{\prime },dx_{T-1}|\overline{\mathbf{x}}_{1:T-2},%
\overline{\mathbf{x}}_{T-1}^{-b_{T-1}},x_{T}^{B_{T}},\overline{\mathbf{a}}%
_{1:T-3},\mathbf{a}_{T-2}^{-b_{t}},\theta \right) .
\end{eqnarray*}
\end{description}
\end{assumption}

Algorithm \ref{MCMCGenStates_algorithm} generates the states using Markov
chain Monte Carlo runs.

\begin{algorithm}[Markov chain Monte Carlo]
\label{MCMCGenStates_algorithm}\ \ 

\begin{description}
\item[Step 1] Run the sequential Monte Carlo algorithm (Algorithm \ref%
{SMC_algorithm}) to obtain $\overline{\mathbf{X}}_{1:T}$ and $\overline{%
\mathbf{A}}_{1:T-1}$.

\item[Step 2] For $t=T,$ sample $B_{T}\sim \mathcal{F}\left( \cdot |\mathbf{W%
}_{T}\right) .$

\item[Step 3] For $t=T-1,$ sample $\tilde{X}_{T-1}^{C_{T-1}}$ as follows

\begin{description}
\item[Step 3.1] compute and normalize the weights%
\begin{equation}
\widetilde{w}_{T-1}^{i}=w_{T-1}^{i}f_{T}\left(
X_{T}^{B^{T}}|X_{T-1}^{i},\theta \right)  \label{MCMCGenStates_5}
\end{equation}%
\begin{equation*}
\widetilde{W}_{T-1}^{i}=\frac{\widetilde{w}_{T-1}^{i}}{\dsum_{j=1}^{N_{T-1}}%
\widetilde{w}_{T-1}^{j}},
\end{equation*}

\item[Step 3.2] sample $B_{T-1}\sim \mathcal{F}\left( \cdot |\widetilde{%
\mathbf{W}}_{T-1}\right) ,$

\item[Step 3.3] set%
\begin{equation*}
\tilde{X}_{T-1}^{\text{old}}=X_{T-1}^{B_{T-1}},
\end{equation*}

\item[Step 3.4] For $j=1,\ldots ,C_{T-1}$

\begin{description}
\item[Step 3.4.1] sample%
\begin{equation*}
\tilde{X}_{T-1}^{j}\sim K_{T-1}\left( \tilde{X}_{T-1}^{\text{old}},\cdot |%
\overline{\mathbf{X}}_{1:T-2},\mathbf{X}_{T-1}^{-B_{T-1}},X_{T}^{B_{T}},%
\overline{\mathbf{A}}_{1:T-3},\mathbf{A}_{T-2}^{-B_{T-1}},\theta \right) ,
\end{equation*}

\item[Step 3.4.2] set%
\begin{equation*}
\tilde{X}_{T-1}^{\text{old}}=\tilde{X}_{T-1}^{j}.
\end{equation*}
\end{description}
\end{description}

\item[Step 4] For $t=T-2,\ldots ,1$

\begin{description}
\item[Step 4.1] compute and normalize the weights%
\begin{equation}
\widetilde{w}_{t}^{i}=w_{t}^{i}f_{t+1}\left( \tilde{X}%
_{t+1}^{C_{t+1}}|X_{t}^{i},\theta \right)  \label{MCMCGenStates_6}
\end{equation}%
\begin{equation*}
\widetilde{W}_{t}^{i}=\frac{\widetilde{w}_{t}^{i}}{\dsum_{j=1}^{N_{t-1}}%
\widetilde{w}_{t}^{j}},
\end{equation*}

\item[Step 4.2] sample $B_{t}\sim \mathcal{F}\left( \cdot |\widetilde{%
\mathbf{W}}_{t}\right) ,$

\item[Step 4.3] set%
\begin{equation*}
\tilde{X}_{t}^{\text{old}}=X_{t}^{B_{t}},
\end{equation*}

\item[Step 4.4] For $j=1,\ldots ,C_{t}$

\begin{description}
\item[Step 4.4.1] sample%
\begin{equation*}
\tilde{X}_{t}^{j}\sim K_{t}\left( \tilde{X}_{t}^{\text{old}},\cdot |%
\overline{\mathbf{X}}_{1:t-1},\mathbf{X}_{t}^{-B_{t}},\tilde{X}_{t+1}^{Ct+1},%
\overline{\mathbf{A}}_{1:t-2},\mathbf{A}_{t-1}^{-B_{t}}\theta \right) ,
\end{equation*}

\item[Step 4.4.2] set%
\begin{equation*}
\tilde{X}_{t}^{\text{old}}=\tilde{X}_{t}^{j}.
\end{equation*}
\end{description}
\end{description}
\end{description}
\end{algorithm}

\subsection{Distributions on the extended space\label{U:Distrib}}

This section first gives the joint probability distribution of the variables
generated by Algorithms \ref{SMC_algorithm} and \ref{MCMCGenStates_algorithm}
before constructing our target distribution and deriving its properties. To
simplify the notation, we group the variables together as $\mathbf{U}%
_{1:T}=\left( \overline{\mathbf{X}}_{1:T},\overline{\mathbf{A}}_{1:T-1},%
\mathbf{B}_{1:T},\widetilde{\mathbf{X}}_{1:T-1}\right) $. We denote the
sample space of $\mathbf{U}_{1:T}$ by%
\begin{equation*}
\mathcal{U}_{1:T}=\dprod_{t=1}^{T}\mathcal{X}^{N_{t}}\times
\dprod_{t=1}^{T-1}\left\{ 1,\ldots ,N_{t}\right\} ^{N_{t+1}}\times
\dprod_{t=1}^{T}\left\{ 1,\ldots ,N_{t}\right\} \times \dprod_{t=1}^{T-1}%
\mathcal{X}^{C_{t}},
\end{equation*}%
and the joint distrbution of $\mathbf{U}_{1:T}$ generated by Algorithms \ref%
{SMC_algorithm} and \ref{MCMCGenStates_algorithm} \ by $\Psi \left( d\mathbf{%
u}_{1:T}\right) =\Psi \left( d\overline{\mathbf{x}}_{1:T},\overline{\mathbf{a%
}}_{1:T-1},\mathbf{b}_{1:T},d\widetilde{\mathbf{x}}_{1:T-1}\right) .$ Let%
\begin{equation*}
r\left( \overline{\mathbf{a}}_{t}|\mathbf{W}_{t}\right) \mathrel{:=}%
\dprod_{i=1}^{N_{t}}\mathcal{F}\left( a_{t}^{i}|\mathbf{W}_{t}\right) .
\end{equation*}%
It is straightforward to show that the distribution $\Psi \left( d\overline{%
\mathbf{x}}_{1:T},\overline{\mathbf{a}}_{1:T-1}\right) $ of the variables $%
\overline{\mathbf{X}}_{1:T},\overline{\mathbf{A}}_{1:T-1}$ generated by
Algorithm \ref{SMC_algorithm} is%
\begin{equation}
\Psi \left( d\overline{\mathbf{x}}_{1:T},\overline{\mathbf{a}}%
_{1:T-1}\right) =\left\{ \dprod_{i=1}^{N_{1}}M_{1}\left(
x_{1}^{i}|y_{1},\theta \right) \right\} \dprod_{t=2}^{T}\left\{ r\left( 
\overline{\mathbf{a}}_{t-1}|\mathbf{W}_{t-1}\right)
\dprod_{i=1}^{N_{t}}M_{t}\left( x_{t}^{i}|y_{t},x_{t-1}^{a_{t-1}^{i}},\theta
\right) \right\} d\overline{\mathbf{x}}_{1:T};  \label{dist_1}
\end{equation}

see \cite{andrieuetal2010} for details.

The conditional distribution $\Psi \left( \mathbf{B}_{1:T},d\widetilde{%
\mathbf{X}}_{1:T-1}|\overline{\mathbf{X}}_{1:T},\overline{\mathbf{A}}%
_{1:T-1}\right) $ generated by Algorithm \ref{MCMCGenStates_algorithm} is

\begin{eqnarray}
&&\Psi \left( \mathbf{b}_{1:T},d\widetilde{\mathbf{x}}_{1:T-1}|\overline{%
\mathbf{x}}_{1:T},\overline{\mathbf{a}}_{1:T-1},\theta \right)  \notag \\
&=&W_{T}^{b_{T}}\widetilde{W}_{T-1}^{b_{T-1}}K_{T-1}\left(
x_{T-1}^{b_{T-1}},d\widetilde{x}_{T-1}^{1}|\overline{\mathbf{x}}_{1:T-2},%
\mathbf{x}_{T-1}^{-b_{T-1}},x_{T}^{b_{T}},\overline{\mathbf{a}}_{1:T-3},%
\mathbf{a}_{T-2}^{-b_{T-1}},\theta \right)  \notag \\
&&\dprod_{j=2}^{C_{T-1}}K_{T-1}\left( \widetilde{x}_{T-1}^{j-1},d\widetilde{x%
}_{T-1}^{j}|\overline{\mathbf{x}}_{1:T-2},\mathbf{x}%
_{T-1}^{-b_{T-1}},x_{T}^{b_{T}},\overline{\mathbf{a}}_{1:T-1},\theta \right)
\notag \\
&&\dprod_{t=1}^{T-2}\left\{ \widetilde{W}_{t}^{b_{t}}K_{t}\left(
x_{t}^{b_{t}},d\widetilde{x}_{t}^{1}|\overline{\mathbf{x}}_{1:t-1},\mathbf{x}%
_{t}^{-b_{t}},\widetilde{x}_{t+1}^{C_{t+1}},\overline{\mathbf{a}},\mathbf{a}%
_{t-1}^{-b_{t}},\theta \right) \right.  \notag \\
&&\left. \dprod_{j=2}^{C_{t}}K_{t}\left( \widetilde{x}_{t}^{j-1},d\widetilde{%
x}_{t}^{j}|\overline{\mathbf{x}}_{1:t-1},\mathbf{x}_{t}^{-b_{t}},\widetilde{x%
}_{t+1}^{C_{t+1}},\overline{\mathbf{a}}_{1:t-2},\mathbf{a}%
_{t-1}^{-b_{t}},\theta \right) \right\} ,  \label{dist_4}
\end{eqnarray}%
and hence the joint distribution of $\Psi \left( d\mathbf{u}_{1:T}\right) $
is the product of (\ref{dist_1}) and (\ref{dist_4}).

We now construct a joint distribution on the variable $U$ that will be the
target distribution of a Markov chain Monte Carlo sampling scheme to
generate a sample from the posterior distribution of the states in a state
space model. To simplify the notation, define%
\begin{equation}
\pi \left( x_{1:T}|\theta \right) \mathrel{:=}p\left( x_{1:T}|y_{1:T},\theta
\right)  \label{dist_5}
\end{equation}%
as the posterior density of the states in the state space model defined by (%
\ref{ssm_1})--(\ref{ssm_4}). The distribution we construct is%
\begin{eqnarray}
&&\Pi \left( d\mathbf{u}_{1:T}|\theta \right) \mathrel{:=}  \notag \\
&&\pi \left( \widetilde{x}_{1}^{C_{1}},\ldots ,\widetilde{x}%
_{T-1}^{C_{T-1}},x_{T}^{b_{T}}|\theta \right) d\widetilde{x}%
_{1}^{C_{1}}\ldots d\widetilde{x}_{T-1}^{C_{T-1}}dx_{T}^{b_{T}}\left( \frac{1%
}{N_{T}}\right)  \notag \\
&&\frac{\Psi \left( d\overline{\mathbf{x}}_{1:T},\overline{\mathbf{a}}%
_{1:T-1}|\theta \right) }{M_{1}\left( dx_{1}^{b_{1}}|y_{1},\theta \right)
\dprod_{t=2}^{T}\left\{ W_{t-1}^{a_{t-1}^{b_{t}}}M_{t}\left(
dx_{t}^{b_{t}}|y_{t},x_{t-1}^{a_{t-1}^{b_{t}}},\theta \right) \right\} } 
\notag \\
&&\dprod_{t=2}^{T}\frac{w_{t-1}^{a_{t-1}^{b_{t}}}f_{t}\left(
x_{t}^{b_{t}}|x_{t-1}^{a_{t-1}^{b_{t}}},\theta \right) }{%
\dsum_{i=1}^{N_{t-1}}w_{t-1}^{i}f_{t}\left( x_{t}^{b_{t}}|x_{t-1}^{i},\theta
\right) }  \notag \\
&&\left( \frac{1}{N_{T-1}}\right) \dprod_{j=2}^{C_{T-1}}K_{T-1}\left( 
\widetilde{x}_{T-1}^{j},d\widetilde{x}_{T-1}^{j-1}|\overline{\mathbf{x}}%
_{1:T-2},\mathbf{x}_{T-1}^{-b_{T-1}},x_{T}^{b_{T}},\overline{\mathbf{a}}%
_{1:T-3},\mathbf{a}_{T-2}^{-b_{T-1}},\theta \right)  \notag \\
&&K_{T-1}\left( \widetilde{x}_{T-1}^{1},dx_{T-1}^{b_{T-1}}|\overline{\mathbf{%
x}}_{1:T-2},\mathbf{x}_{T-1}^{-b_{T-1}},x_{T}^{b_{T}},\overline{\mathbf{a}}%
_{1:T-3},\mathbf{a}_{T-2}^{-b_{T-1}},\theta \right)  \notag \\
&&\dprod_{t=1}^{T-2}\left\{ \left( \frac{1}{N_{t}}\right)
\dprod_{j=2}^{C_{t}}K_{t}\left( \widetilde{x}_{t}^{j},d\widetilde{x}%
_{t}^{j-1}|\overline{\mathbf{x}}_{1:t-1},\mathbf{x}_{t}^{-b_{t}},\widetilde{x%
}_{t+1}^{Ct+1},\overline{\mathbf{a}}_{1:t-2},\mathbf{a}_{t-1}^{-b_{t}},%
\theta \right) \right.  \notag \\
&&\left. K_{t}\left( \widetilde{x}_{t}^{1},dx_{t}^{b_{t}}|\overline{\mathbf{x%
}}_{1:t-1},\mathbf{x}_{t}^{-b_{t}},\widetilde{x}_{t+1}^{Ct+1},\overline{%
\mathbf{a}}_{1:t-2},\mathbf{a}_{t-1}^{-b_{t}},\theta \right) \right\} ,
\label{dist_6}
\end{eqnarray}%
which is well defined by Assumption \ref{assumption:support}.

The following lemma describes the properties of the distribution defined in (%
\ref{dist_6}). Its proof is given in the Appendix.

\begin{lemma}

\label{MHRatio_lemma}

\begin{description}
\item[(i)] The joint distribution $\Pi \left( d\mathbf{u}_{1:T}|\theta
\right) $ has marginal distribution%
\begin{equation*}
\Pi \left( d\widetilde{x}_{1}^{C_{1}},\ldots ,d\widetilde{x}%
_{T-1}^{C_{T-1}},dx_{T}^{b_{T}},b_{T}|\theta \right) =\pi \left( \widetilde{x%
}_{1}^{C_{1}},\ldots ,\widetilde{x}_{T-1}^{C_{T-1}},x_{T}^{b_{T}}|\theta
\right) d\widetilde{x}_{1}^{C_{1}}\ldots ,d\widetilde{x}%
_{T-1}^{C_{T-1}}dx_{T}^{b_{T}}\left( \frac{1}{N_{T}}\right) .
\end{equation*}

\item[(ii)] For all $\theta \in \Theta ,$ the measures $\Pi \left( \cdot
|\theta \right) $ and $\Psi \left( \cdot |\theta \right) $ are equivalent.

\item[(iii)] There exists a version of the density%
\begin{equation*}
h\left( \mathbf{u}_{1:T}|\theta \right) =\frac{\Pi \left( d\mathbf{u}%
_{1:T}|\theta \right) }{\Psi \left( d\mathbf{u}_{1:T}|\theta \right) }
\end{equation*}%
with%
\begin{equation}
h\left( \mathbf{u}_{1:T}|\theta \right) =\frac{\dprod_{t=1}^{T}\left\{
\left( \dsum_{i=1}^{N_{t}}w_{t}^{i}\right) \left( \frac{1}{N_{t}}\right)
\right\} }{p\left( \mathbf{y}_{1:T}|\theta \right) }  \label{dist_8}
\end{equation}
\end{description}
\end{lemma}

Lemma \ref{PG_lemma_2} shows how to generate a sample from the distribution%
\begin{equation}
\Pi \left( \mathbf{b}_{1:T},d\widetilde{\mathbf{x}}_{1:T-1}|\overline{%
\mathbf{x}}_{1:T},\overline{\mathbf{a}}_{1:T-1},\theta \right) .
\label{PG_2}
\end{equation}%
Its proof is given in the appendix.

\begin{lemma}
\label{PG_lemma_2}%
\begin{equation*}
\Pi \left( \mathbf{b}_{1:T},d\widetilde{\mathbf{x}}_{1:T-1}|\overline{%
\mathbf{x}}_{1:T},\overline{\mathbf{a}}_{1:T-1},\theta \right) =\Psi \left( 
\mathbf{b}_{1:T},d\widetilde{\mathbf{x}}_{1:T-1}|\overline{\mathbf{x}}_{1:T},%
\overline{\mathbf{a}}_{1:T-1},\theta \right) ,
\end{equation*}

and hence Algorithm \ref{MCMCGenStates_algorithm} generates a sample from
the distribution given in (\ref{PG_2}).
\end{lemma}

\subsection{Conditional sequential Monte Carlo\label{U:PG}}

This section gives a conditional sequential Monte Carlo algorithm that is
used to construct a particle Gibbs step later in the paper. We first
describe the algorithm and derive its properties. Section \ref{U:Estimation}
shows how to use it in Markov chain Monte Carlo sampling schemes.

Algorithm \ref{PG_algorithm} generates from the conditional distribution%
\begin{equation}
\Pi \left( d\overline{\mathbf{x}}_{1:T-1},d\mathbf{x}_{T}^{-b_{T}},\overline{%
\mathbf{a}}_{1:T-1},\mathbf{b}_{1:T-1},d\widetilde{\mathbf{x}}%
_{1}^{-C_{1}},\ldots ,d\widetilde{\mathbf{x}}_{T-1}^{-C_{T-1}}|\widetilde{x}%
_{1}^{C_{1}},\ldots ,\widetilde{x}_{T-1}^{C_{T-1}},x_{T}^{b_{T}},b_{T},%
\theta \right) .  \label{PG_1}
\end{equation}

\begin{algorithm}[Particle Gibbs]
\label{PG_algorithm}\ \ 

\begin{description}
\item[Step 1] For $t=1,\ldots ,T-1$

\begin{description}
\item[Step 1.1] generate $B_{t}\sim \limfunc{Uniform}\left\{ 1,\ldots
N_{t}\right\} ,$

\item[Step 1.2] generate $A_{t-1}^{-B_{t}}$ and $X_{t}^{-B_{t}}$ using the
Sequential Monte Carlo Algorithm \ref{SMC_algorithm},

\item[Step 1.3] set $\widetilde{X}_{t}^{\text{old}}=\widetilde{x}%
_{t}^{C_{t}},$

\item[Step 1.4] for $j=C_{t}-1,\ldots ,1$

\begin{description}
\item[Step 1.4.1] generate $\widetilde{X}_{t}^{j}$ from%
\begin{equation*}
K_{t}\left( \tilde{X}_{t}^{\text{old}},\cdot |\overline{\mathbf{X}}_{1:t-1},%
\mathbf{X}_{t}^{-B_{t}},\tilde{X}_{t+1}^{C_{t+1}},\overline{\mathbf{A}}%
_{1:t-2},\mathbf{A}_{t-1}^{-B_{t}},\theta \right) ,
\end{equation*}

\item[Step 1.4.2] set $\widetilde{X}_{t}^{\text{old}}=\widetilde{X}_{t}^{j},$
\end{description}

\item[Step 1.5] generate $X_{t}^{B_{t}}$ from%
\begin{equation*}
K_{t}\left( \tilde{X}_{t}^{\text{old}},\cdot |\overline{\mathbf{X}}_{1:t-1},,%
\mathbf{X}_{t}^{-B_{t}},\tilde{X}_{t+1}^{C_{t+1}},\overline{\mathbf{A}}%
_{1:t-2},\mathbf{A}_{t-1}^{-B_{t}},\theta \right) ,
\end{equation*}

\item[Step 1.6] if $t>1$ then generate $\mathbf{A}_{t-1}^{B_{t}}$ as follows

\begin{description}
\item[Step 1.6.1] compute and normalize the weights%
\begin{equation*}
v_{t-1}^{i}=w_{t-1}^{i}f_{t}\left( X_{t}^{B_{t}}|X_{t-1}^{i},\theta \right)
\end{equation*}%
\begin{equation*}
V_{t-1}^{i}=\frac{v_{t-1}^{i}}{\dsum_{j=1}^{N_{t-1}}v_{t}^{j}},
\end{equation*}

\item[Step 1.6.2] generate $\mathbf{A}_{t-1}^{B_{t}}\sim \mathcal{F}\left(
\cdot |\mathbf{V}_{t-1}\right) .$
\end{description}
\end{description}

\item[Step 2] For $t=T,$ generate $\mathbf{A}_{T-1}^{B_{T}}$ as follows

\begin{description}
\item[Step 2.1] compute and normalize the weights%
\begin{equation*}
v_{T-1}^{i}=w_{T-1}^{i}f_{T}\left( X_{T}^{B_{T}}|X_{T-1}^{i},\theta \right)
\end{equation*}%
\begin{equation*}
V_{T-1}^{i}=\frac{v_{T-1}^{i}}{\dsum_{j=1}^{N_{T-1}}v_{T-1}^{j}},
\end{equation*}

\item[Step 2.2] generate $\mathbf{A}_{T-1}^{B_{T}}\sim \mathcal{F}\left(
\cdot |\mathbf{V}_{T-1}\right) $.
\end{description}

\item[Step 3] For $t=T,$ generate $\mathbf{A}_{T-1}^{-B_{T}}$ and $\mathbf{X}%
_{T}^{-B_{T}}$ using the Sequential Monte Carlo Algorithm \ref{SMC_algorithm}%
.
\end{description}
\end{algorithm}

Lemma \ref{PGAlgorithm_lemma_1} gives the properties of the algorithm
described above. Its proof is given in the Appendix.

\begin{lemma}
\label{PGAlgorithm_lemma_1}Algorithm \ref{PG_algorithm} generates a sample
from the distribution given in (\ref{PG_1}).
\end{lemma}

\section{Estimation for State Space Models\label{U:Estimation}}

This section shows how to use the algorithms and distributions in Section %
\ref{U:genstates} to carry out smoothing and inference for state space
models. We first consider the smoothing case and then consider several
approaches to parameter estimation. We also consider ergodicity results for
the methods.

\subsection{Smoothing approaches \label{U:PIMH}}

The simplest application of the results in Section \ref{U:genstates} is the
smoothing problem where the parameter $\theta $ is regarded as fixed and
known and we wish to generate a sample from the density $\pi \left(
x_{1:T}|\theta \right) $ defined in (\ref{dist_5}). There are several
possible smoothing approaches. We first describe a particle independent
Metropolis-Hastings approach using the following sampling scheme that
describes one sweep of a Markov chain Monte Carlo algorithm.

\begin{samplingscheme}[PMMH Smoothing]
\label{PIMH_sscheme}Given $\mathbf{U}_{1:T}$ and $\theta $

\begin{description}
\item[Step 1] Sample $\mathbf{U}_{1:T}^{\prime }\sim \Psi \left( \cdot
|\theta \right) $ using Algorithms \ref{SMC_algorithm} and \ref%
{MCMCGenStates_algorithm}.

\item[Step 2] Accept $\mathbf{U}_{1:T}^{\prime }$ with probability 
\begin{equation}
\min \left[ 1,\frac{\dprod_{t=1}^{T}\left\{ \dsum_{i=1}^{N_{t}}\left(
w_{t}^{i}\right) ^{^{\prime }}\right\} }{\dprod_{t=1}^{T}\left\{ \left(
\dsum_{i=1}^{N_{t}}w_{t}^{i}\right) \right\} }\right] .  \label{pimh_1}
\end{equation}
\end{description}
\end{samplingscheme}

\begin{remark}
The acceptance probability (\ref{pimh_1}) only requires the output from
Algorithm \ref{SMC_algorithm}, so it is possible to run Algorithm \ref%
{SMC_algorithm} in Step 1 and only run Algorithm \ref%
{MCMCGenStates_algorithm} if the Metropolis-Hasting proposal is accepted in
Step 2.
\end{remark}

\begin{remark}
The acceptance probability (\ref{pimh_1}) is the same expression as obtained
for the particle independent Metropolis-Hasting methods described in \cite%
{andrieuetal2010} and \cite{olssonryden2011}. The advantage of our method is
that Algorithm \ref{MCMCGenStates_algorithm} generates new values of the
states that are not restricted to the values from the sequential Monte Carlo
output from Algorithm \ref{SMC_algorithm}.
\end{remark}

It is also possible to use a particle Gibbs sampler to generate a sample
from the density $\pi \left( x_{1:T}|\theta \right) $. We use the following
sampling scheme that describes one sweep of a Markov chain Monte Carlo
algorithm.

\begin{samplingscheme}[Particle Gibbs Smoothing]
\label{PGsmoothing_sscheme}Given $\widetilde{X}_{1}^{C_{1}},\ldots ,%
\widetilde{X}_{T-1}^{C_{T-1}},X_{T}^{B_{T}},B_{T},\theta $

\begin{description}
\item[Step 1] Run the particle Gibbs Algorithm \ref{PG_algorithm} to sample%
\begin{equation*}
\overline{\mathbf{X}}_{1:T-1},\overline{\mathbf{X}}_{T}^{-B_{T}},\overline{%
\mathbf{A}}_{1:T-1},\mathbf{B}_{1:T-1},\widetilde{\mathbf{X}}%
_{1}^{-C_{1}},\ldots ,\widetilde{\mathbf{X}}_{T-1}^{-C_{T-1}}
\end{equation*}%
from%
\begin{equation*}
\Pi \left\{ d\overline{\mathbf{X}}_{1:T-1},d\overline{\mathbf{X}}%
_{T}^{-B_{T}},\overline{\mathbf{A}}_{1:T-1},\mathbf{B}_{1:T-1},d\widetilde{%
\mathbf{X}}_{1}^{-C_{1}},\ldots ,d\widetilde{\mathbf{X}}_{T-1}^{-C_{T-1}}|%
\widetilde{X}_{1}^{C_{1}},\ldots ,\widetilde{X}%
_{T-1}^{C_{T-1}},X_{T}^{B_{T}},B_{T},\theta \right\} .
\end{equation*}

\item[Step 2] Sample $\mathbf{B}_{1:T},\widetilde{\mathbf{X}}_{1:T-1}$ from%
\begin{equation*}
\Pi \left( \mathbf{B}_{1:T},d\widetilde{\mathbf{X}}_{1:T-1}|\overline{%
\mathbf{X}}_{1:T},\overline{\mathbf{A}}_{1:T-1},\theta \right) ,
\end{equation*}%
using Algorithm \ref{MCMCGenStates_algorithm}.
\end{description}
\end{samplingscheme}

The potential advantage of Sampling Scheme \ref{PGsmoothing_sscheme} is that
it is a Gibbs sampler and hence avoids the inefficiency involved in
rejecting proposed moves. A potential disadvantage of Sampling Scheme \ref%
{PGsmoothing_sscheme} is that it is computationally more expensive than
Sampling Scheme \ref{PIMH_sscheme}.

\begin{remark}
A random scan version of Sampling Scheme \ref{PGsmoothing_sscheme} may be
more efficient since Step 1 can be sampled with low probability to avoid the
computational cost of running the particle Gibbs Algorithm \ref{PG_algorithm}
for each iterate.
\end{remark}

\subsection{Inference using general sampling schemes\label{U:Hybrid}}

This section considers full Bayesian inference where both the auxilary
variables and the parameters are generated. There are three possible
approaches to generating the parameters: particle Marginal
Metropolis-Hastings, particle Gibbs and particle Metropolis within Gibbs
steps. We illustrate the method with an example where the parameters are
partitioned into $p$ components $\theta =(\theta _{1},\ldots ,\theta _{p}),$
where each component may be a vector. Let $\Theta =\Theta _{1}\times \ldots
\times \Theta _{p}$ be the corresponding partition of the parameter space.
We will use the notation%
\begin{equation*}
\theta _{-i}=(\theta _{1},\ldots ,\theta _{i-1},\theta _{i+1},\ldots ,\theta
_{p})\text{.}
\end{equation*}%
We denote the prior density for $\theta $ by $p\left( \theta \right) $ and
the posterior density by $\pi \left( \theta \right) $ and we assume that
both densities are strictly positive.

Let $0\leq p_{1}\leq p$. The following sampling scheme generates the
parameters $\theta _{1},\ldots ,\theta _{p_{1}}$ using particle Marginal
Metropolis-Hasting steps and the parameters $\theta _{p_{1}+1},\ldots
,\theta _{p}$ using particle Gibbs or particle Metropolis within Gibbs
steps. We call this a general sampler as described in \cite{mendesgen2014}.

\begin{samplingscheme}
\label{Hybrid_sscheme}Given $\mathbf{U}_{1:T}$ and $\theta $

\begin{description}
\item[Part 1] (PMMH sampling) For $i=1,\ldots ,p_{1}$

Step $i$:

\begin{description}
\item[(a)] sample%
\begin{equation}
\theta _{i}^{\prime }\sim q_{i}\left( \cdot |\mathbf{U}_{1:T},\theta
_{-i},\theta _{i}\right)  \label{hybrid_metprop1}
\end{equation}

\item[(b)] sample 
\begin{equation*}
\mathbf{U}_{1:T}^{\prime }\sim \Psi \left( \cdot |\theta _{-i},\theta
_{i}^{\prime }\right)
\end{equation*}%
using Algorithms \ref{SMC_algorithm} and \ref{MCMCGenStates_algorithm}.

\item[(c)] accept $\mathbf{U}_{1:T}^{\prime },\theta _{i}^{\prime }$ with
probability equal to%
\begin{equation}
1\wedge \frac{\dprod_{t=1}^{T}\left\{ \dsum_{i=1}^{N_{t}}\left(
w_{t}^{i}\right) ^{^{\prime }}\right\} p\left( \theta _{i}^{\prime }|\theta
_{-i}\right) q_{i}\left( \theta _{i}|\mathbf{U}_{1:T}^{\prime },\theta
_{-i},\theta _{i}^{\prime }\right) }{\dprod_{t=1}^{T}\left\{ \left(
\dsum_{i=1}^{N_{t}}w_{t}^{i}\right) \right\} p\left( \theta _{i}|\theta
_{-i}\right) q_{i}\left( \theta _{i}^{\prime }|\mathbf{U}_{1:T},\theta
_{-i},\theta _{i}\right) }.  \label{hybrid_accprob1}
\end{equation}
\end{description}

\item[Part 2] (PG or PMwG sampling) For $i=p_{1}+1,\ldots ,p$

Step $i$:

\begin{description}
\item[(a)] sample $\theta _{i}^{\prime }\sim q_{i}(\cdot |\widetilde{X}%
_{1}^{C_{1}},\ldots ,\widetilde{X}_{T-1}^{C_{T-1}},X_{T}^{B_{T}},B_{T},%
\theta _{-i},\theta _{i}),$

\item[(b)] run the Particle Gibbs Algorithm \ref{PG_algorithm} to sample%
\begin{equation*}
\overline{\mathbf{X}}_{1:T-1}^{\prime },\left( \mathbf{X}_{T}^{-b_{T}}%
\right) ^{\prime },\overline{\mathbf{A}}_{1:T-1}^{\prime },\mathbf{B}%
_{1:T-1}^{\prime },\left( \widetilde{\mathbf{X}}_{1}^{-C_{t}}\right)
^{\prime },\ldots ,\left( \widetilde{\mathbf{X}}_{T-1}^{-C_{T-1}}\right)
^{\prime }
\end{equation*}%
from%
\begin{equation*}
\Pi \left( d\overline{\mathbf{X}}_{1:T-1},d\mathbf{X}_{T}^{-B_{T}},\overline{%
\mathbf{A}}_{1:T-1},\mathbf{B}_{1:T-1},d\widetilde{\mathbf{X}}%
_{1}^{-C_{1}},\ldots ,d\widetilde{\mathbf{X}}_{T-1}^{-C_{T-1}}|\widetilde{X}%
_{1}^{C_{1}},\ldots ,\widetilde{X}_{T-1}^{C_{T-1}},X_{T}^{B_{T}},B_{T},%
\theta _{-i},\theta _{i}^{\prime }\right) ,
\end{equation*}

\item[(c)] accept the proposed values $\overline{\mathbf{X}}_{1:T-1}^{\prime
},\left( \mathbf{X}_{T}^{-b_{T}}\right) ^{\prime },\overline{\mathbf{A}}%
_{1:T-1}^{\prime },\mathbf{B}_{1:T-1}^{\prime },\left( \widetilde{\mathbf{X}}%
_{1}^{-C_{t}}\right) ^{\prime },\ldots ,\left( \widetilde{\mathbf{X}}%
_{T-1}^{-C_{T-1}}\right) ^{\prime }$ and $\theta _{i}^{\prime }$ with
probability 
\begin{equation}
1\wedge \frac{\pi \left( \theta _{i}^{\prime }|\widetilde{X}%
_{1}^{C_{1}},\ldots ,\widetilde{X}_{T-1}^{C_{T-1}},X_{T}^{B_{T}},B_{T},%
\theta _{-i}\right) }{\pi \left( \theta _{i}|\widetilde{X}%
_{1}^{C_{1}},\ldots ,\widetilde{X}_{T-1}^{C_{T-1}},X_{T}^{B_{T}},B_{T},%
\theta _{-i}\right) }\frac{q_{i}(\theta _{i}|\widetilde{X}%
_{1}^{C_{1}},\ldots ,\widetilde{X}_{T-1}^{C_{T-1}},X_{T}^{B_{T}},B_{T},%
\theta _{-i},\theta _{i}^{\prime })}{q_{i}(\theta _{i}^{\prime }|\widetilde{X%
}_{1}^{C_{1}},\ldots ,\widetilde{X}_{T-1}^{C_{T-1}},X_{T}^{B_{T}},B_{T},%
\theta _{-i},\theta _{i})},  \notag
\end{equation}

\item[(d)] sample $\mathbf{B}_{1:T},\widetilde{\mathbf{X}}_{1:T-1}$ from%
\begin{equation*}
\Pi \left( \mathbf{B}_{1:T},d\widetilde{\mathbf{X}}_{1:T-1}|\overline{%
\mathbf{X}}_{1:T},\overline{\mathbf{A}}_{1:T-1},\theta \right) .
\end{equation*}
\end{description}
\end{description}
\end{samplingscheme}

We will assume that the Metropolis-Hasting proposals $q_{i}\left( \cdot |%
\mathbf{U}_{1:T},\theta _{-i},\theta _{i}\right) $ for $i=1,\ldots ,p_{1}$
and $q_{i}(\cdot |\widetilde{X}_{1}^{C_{1}},\ldots ,\widetilde{X}%
_{T-1}^{C_{T-1}},X_{T}^{B_{T}},B_{T},\theta _{-i},\theta _{i})$ for $%
i=p_{1}+1,\ldots ,p$ are strictly positive densities.

\begin{remark}
A similar sampling scheme is given in \cite{mendesgen2014} for the augmented
models from \cite{andrieuetal2010} and \cite{olssonryden2011}. \cite%
{mendesgen2014} show that more general sampling schemes perform better than
the particle Marginal Metropolis-Hastings approach in \cite{andrieuetal2010}
and the particle Gibbs approach in \cite{lindstenetal2014} for some
applications.
\end{remark}

\begin{remark}
Similar comments apply between Sampling Scheme \ref{PIMH_sscheme} and Part 1
of Sampling Scheme \ref{Hybrid_sscheme}. Similarly to Sampling Scheme \ref%
{PIMH_sscheme}, only the terms in the acceptance probabilities (\ref%
{hybrid_accprob1}) are required in Part 1, Step $i(a)$. We have shown the
most general case for the Metropolis-Hastings proposal (\ref{hybrid_metprop1}%
) which requires both Algorithms \ref{SMC_algorithm} and \ref%
{MCMCGenStates_algorithm} to be run. Simpler Metropolis-Hasting proposals of
the form%
\begin{equation*}
q_{i}\left( \cdot |\overline{\mathbf{X}}_{1:T},\overline{\mathbf{A}}%
_{1:T-1},\theta \right)
\end{equation*}%
would only require Algorithm \ref{MCMCGenStates_algorithm} to be run if the
Metropolis-Hasting proposal is accepted in Part 1 Step $i$(c).
\end{remark}

\begin{remark}
In Part 2 of Sampling Scheme \ref{Hybrid_sscheme}, it is only necessary to
generate the values in Step $i\left( b\right) $ if the Metropolis-Hasting
proposal is accepted in Step $i\left( c\right) $.
\end{remark}

\begin{remark}
In Part 2 of Sampling Scheme \ref{Hybrid_sscheme}, it is possible to remove
Step $i\left( d\right) $ for $i=p_{1}+1,\ldots ,p-1$.
\end{remark}

\subsection{Ergodicity\label{U:Ergodicity}}

This section gives sufficient conditions for Sampling Schemes \ref%
{PIMH_sscheme} to \ref{Hybrid_sscheme} to converge to their stationary
distributions in total variation norm and uniform convergence.

Note that, by construction, Sampling Schemes \ref{PIMH_sscheme} to \ref%
{Hybrid_sscheme} have correct invariant distributions, so to prove
convergence in total variation norm it is sufficient to prove that the
corresponding Markov chains are irreducible and aperiodic and then use
standard Markov chain convergence results -- see, for example, Theorem 4 in 
\cite{robertsrosenthal2004}.

Theorem \ref{PIMH_theorem_1} gives sufficient conditions for Sampling Scheme %
\ref{PIMH_sscheme} to converge to $\Pi $ in total variation norm.

\begin{theorem}
\label{PIMH_theorem_1} If Assumptions \ref{assumption:support} and \ref%
{assumption:detailedbalance} hold, then for all $\theta \in \Theta $
Sampling Scheme \ref{PIMH_sscheme} generates a sequence\ $\left\{ \mathbf{U}%
_{1:T}\left( j\right) \right\} $, whose distributions $\left\{ \mathcal{L}%
\left\{ \mathbf{U}_{1:T}\left( j\right) \in \cdot \right\} \right\} $ satisfy%
\begin{equation*}
\left\vert \mathcal{L}\left\{ \mathbf{U}_{1:T}\left( j\right) \in \cdot
\right\} -\Pi \left( \cdot |\theta \right) \right\vert _{TV}\rightarrow 0%
\text{\qquad as }j\rightarrow \infty \text{.}
\end{equation*}
\end{theorem}

Its proof follows directly from Lemma \ref{MHRatio_lemma} and standard
convergence results for Markov chains. Part (iii) of Lemma \ref%
{MHRatio_lemma} shows that Sampling Scheme \ref{PIMH_sscheme} is an
independent Metropolis-Hasting algorithm with target distribution $\Pi
\left( \cdot \right) $ and Part (ii) of Lemma \ref{MHRatio_lemma} shows that
the Markov chain is irreducible and aperiodic.

Theorem \ref{PGsmoothing_theorem_1} uses Assumption \ref%
{assumption:productkernel} to give sufficient conditions for the convergence
in total variation norm of Sampling Scheme \ref{PGsmoothing_sscheme}.

\begin{assumption}
\label{assumption:productkernel}For $t=1,\ldots ,T-1,$ the product kernel%
\begin{equation*}
K_{t}^{C_{t}}\left( x_{t},dx_{t}^{\prime }|\overline{\mathbf{x}}_{1:t-1},%
\mathbf{x}_{t}^{-b_{t}},\widetilde{x}_{t+1}^{Ct+1},\overline{\mathbf{a}}%
_{1:t-2},\mathbf{a}_{t-1}^{-b_{t}},\theta \right) \gg dx_{t}^{\prime }
\end{equation*}
for all values of $x_{t},\overline{\mathbf{x}}_{1:t-1},\mathbf{x}%
_{t}^{-b_{t}},\widetilde{x}_{t+1}^{Ct+1},\overline{\mathbf{a}}_{1:t-2},%
\mathbf{a}_{t-1}^{-b_{t}}$ and all $\theta \in \Theta $.
\end{assumption}

\begin{theorem}
\label{PGsmoothing_theorem_1} If Assumptions \ref{assumption:support} to \ref%
{assumption:productkernel} hold, then for all $\theta \in \Theta $ Sampling
Scheme \ref{PGsmoothing_sscheme} generates a sequence\ $\left\{ \mathbf{U}%
_{1:T}\left( j\right) \right\} $, whose distributions $\left\{ \mathcal{L}%
\left\{ \mathbf{U}_{1:T}\left( j\right) \in \cdot \right\} \right\} $ satisfy%
\begin{equation*}
\left\vert \mathcal{L}\left\{ \mathbf{U}_{1:T}\left( j\right) \in \cdot
\right\} -\Pi \left( \cdot |\theta \right) \right\vert _{TV}\rightarrow 0%
\text{\qquad as }j\rightarrow \infty \text{.}
\end{equation*}
\end{theorem}

The proof is given in the appendix.

Theorem \ref{Hybrid_sscheme_theorem_1} gives sufficient conditions for the
convergence in total variation norm of Sampling Scheme \ref{Hybrid_sscheme}.
We use the following notation. Let $\left\{ V\left( n\right) ,n=1,2,\ldots
\right\} $ denote the iterates of the Markov chains on the state space $%
\mathcal{V}=\mathcal{U}_{1:T}\times \Theta $.

\begin{theorem}
\label{Hybrid_sscheme_theorem_1}If Assumptions \ref{assumption:support} to %
\ref{assumption:productkernel} hold, then Sampling Scheme \ref%
{Hybrid_sscheme} generates a sequence\ $\left\{ \mathbf{V}\left( j\right)
\right\} $, whose distributions $\left\{ \mathcal{L}\left\{ \mathbf{V}\left(
j\right) \in \cdot \right\} \right\} $ satisfy%
\begin{equation*}
\left\vert \mathcal{L}\left\{ \mathbf{V}\left( j\right) \in \cdot \right\}
-\Pi \left( \cdot \right) \right\vert _{TV}\rightarrow 0\text{\qquad as }%
j\rightarrow \infty \text{.}
\end{equation*}
\end{theorem}

The proof is given in the Appendix.

To derive results on uniform convergence we use a similar approach to \cite%
{andrieuroberts2009}, \cite{andrieuvihola2012}, \cite{lindstenschon2012} and 
\cite{mendesgen2014} and relate Sampling Scheme \ref{Hybrid_sscheme} to the 
\textit{ideal} sampling scheme defined below. The extension from \cite%
{mendesgen2014} to Sampling Scheme \ref{Hybrid_sscheme} is straightforward,
but we include the results for completeness. Similarly to \cite%
{mendesgen2014}, we define the \textit{ideal} sampling scheme:

\begin{samplingscheme}
\label{Ideal_hybrid_sscheme}Given $\mathbf{U}_{1:T}$ and $\theta $

\begin{description}
\item[Part 1] (PMMH sampling) For $i=1,\ldots ,p_{1}$

Step $i$:

\begin{description}
\item[(a)] sample%
\begin{equation}
\theta _{i}^{\prime }\sim q_{i}\left( \cdot |\mathbf{U}_{1:T},\theta
_{-i}\right) ,  \label{ideal_hybrid_metprop1}
\end{equation}

\item[(b)] sample $\mathbf{U}_{1:T}^{\prime }$ from%
\begin{equation*}
\Pi \left( \cdot |\theta _{-i},\theta _{i}^{\prime }\right) ,
\end{equation*}

\item[(c)] accept $\mathbf{U}_{1:T}^{\prime },\theta _{i}^{\prime }$ with
probability equal to%
\begin{equation}
1\wedge \frac{\pi \left( \theta _{i}^{\prime }|\theta _{-i}\right)
q_{i}\left( \theta _{i}|\mathbf{U}_{1:T}^{\prime },\theta _{-i},\theta
_{i}^{\prime }\right) }{\pi \left( \theta _{i}|\theta _{-i}\right)
q_{i}\left( \theta _{i}^{\prime }|\mathbf{U}_{1:T},\theta _{-i},\theta
_{i}\right) }.  \label{ideal_hybrid_accprob1}
\end{equation}
\end{description}

\item[Part 2] (PG or PMwG sampling) For $i=p_{1}+1,\ldots ,p$

Step $i$:

\begin{description}
\item[(a)] sample 
\begin{equation}
\theta _{i}^{\prime }\sim q_{i}(\cdot |\widetilde{X}_{1}^{C_{1}},\ldots ,%
\widetilde{X}_{T-1}^{C_{T-1}},X_{T}^{B_{T}},B_{T},\theta _{-i},\theta _{i}),
\label{ideal_hybrid_metprop2}
\end{equation}

\item[(b)] run the Particle Gibbs Algorithm \ref{PG_algorithm} to sample%
\begin{equation*}
\overline{\mathbf{X}}_{1:T-1}^{\prime },\left( \mathbf{X}_{T}^{-b_{T}}%
\right) ^{\prime },\overline{\mathbf{A}}_{1:T-1}^{\prime },\mathbf{B}%
_{1:T-1}^{\prime },\left( \widetilde{\mathbf{X}}_{1}^{-C_{t}}\right)
^{\prime },\ldots ,\left( \widetilde{\mathbf{X}}_{T-1}^{-C_{T-1}}\right)
^{\prime }
\end{equation*}%
from%
\begin{equation*}
\Pi \left\{ d\overline{\mathbf{X}}_{1:T-1},d\mathbf{X}_{T}^{-B_{T}},%
\overline{\mathbf{A}}_{1:T-1},\mathbf{B}_{1:T},d\widetilde{\mathbf{X}}%
_{1}^{-C_{1}},\ldots ,d\widetilde{\mathbf{X}}_{T-1}^{-C_{T-1}}|\widetilde{X}%
_{1}^{C_{1}},\ldots ,\widetilde{X}_{T-1}^{C_{T-1}},X_{T}^{B_{T}},B_{T},%
\theta _{-i},\theta _{i}^{\prime }\right\} ,
\end{equation*}

\item[(c)] accept the proposed values $\overline{\mathbf{X}}_{1:T-1}^{\prime
},\left( \mathbf{X}_{T}^{-b_{T}}\right) ^{\prime },\overline{\mathbf{A}}%
_{1:T-1}^{\prime },\mathbf{B}_{1:T-1}^{\prime },\left( \widetilde{\mathbf{X}}%
_{1}^{-C_{t}}\right) ^{\prime },\ldots ,\left( \widetilde{\mathbf{X}}%
_{T-1}^{-C_{T-1}}\right) ^{\prime }$ and $\theta _{i}^{\prime }$ with
probability 
\begin{equation}
1\wedge \frac{\pi \left( \theta _{i}^{\prime }|\widetilde{X}%
_{1}^{C_{1}},\ldots ,\widetilde{X}_{T-1}^{C_{T-1}},X_{T}^{B_{T}},B_{T},%
\theta _{-i}\right) }{\pi \left( \theta _{i}|\widetilde{X}%
_{1}^{C_{1}},\ldots ,\widetilde{X}_{T-1}^{C_{T-1}},X_{T}^{B_{T}},B_{T},%
\theta _{-i}\right) }\frac{q_{i}(\theta _{i}|\widetilde{X}%
_{1}^{C_{1}},\ldots ,\widetilde{X}_{T-1}^{C_{T-1}},X_{T}^{B_{T}},B_{T},%
\theta _{-i},\theta _{i}^{\prime })}{q_{i}(\theta _{i}^{\prime }|\widetilde{X%
}_{1}^{C_{1}},\ldots ,\widetilde{X}_{T-1}^{C_{T-1}},X_{T}^{B_{T}},B_{T},%
\theta _{-i},\theta _{i})},  \notag
\end{equation}

\item[(d)] sample $\mathbf{B}_{1:T},\widetilde{\mathbf{X}}_{1:T-1}$ from%
\begin{equation*}
\Pi \left( \mathbf{B}_{1:T},d\widetilde{\mathbf{X}}_{1:T-1}|\overline{%
\mathbf{X}}_{1:T},\overline{\mathbf{A}}_{1:T-1},\theta \right) .
\end{equation*}
\end{description}
\end{description}
\end{samplingscheme}

We call Sampling Scheme \ref{Ideal_hybrid_sscheme} an \textit{ideal}
sampling scheme because in Step $i$(b) \ of Part 1 we sample the auxiliary
variables $\mathbf{U}_{1:T}^{\prime }$ from their conditional distribution $%
\Pi \left( \cdot |\theta \right) $, whereas Sampling Scheme \ref%
{Hybrid_sscheme} Step $i$(b) of Part 1 uses the Metropolis-Hasting proposal $%
\Psi \left( \cdot |\theta \right) $. Thus, comparing the two Sampling
Schemes allows us to study the effect of this Metropolis-Hastings proposal
on the convergence of the sampler. Let $P\left( v;\cdot \right) $ be the
substochastic transition kernel of Sampling Scheme \ref{Hybrid_sscheme} that
defines the probabilities for accepted Metropolis-Hastings moves and let $%
\widetilde{P}\left( v;\cdot \right) $ be the corresponding substochastic
kernel for Sampling Scheme \ref{Ideal_hybrid_sscheme}. The following theorem
gives sufficient conditions for the existence of minorization conditions for
Sampling Scheme \ref{Hybrid_sscheme}, which, from \cite{robertsrosenthal2004}
are equivalent to uniform ergodicity.

\begin{theorem}
\label{Hybrid_sscheme_theorem_2} Suppose that

\begin{description}
\item[(i)] Sampling Scheme \ref{Ideal_hybrid_sscheme} satisfies the
following minorization condition:\ there exists a constant $\varepsilon >0$,
a number $n_{0}\geq 1$, and a probability measure $\nu $ on $\mathcal{V}$
such that $\widetilde{P}^{n_{0}}(v;A)\geq \epsilon \,\nu (A)$ for all $v\in 
\mathcal{V},A\in \mathcal{B}\left( \mathcal{V}\right) $.

\item[(ii)] 
\begin{equation*}
h\left( \mathbf{u}_{1:T}|\theta \right) =\frac{\Pi \left( d\mathbf{u}%
_{1:T}|\theta \right) }{\Psi \left( d\mathbf{u}_{1:T}|\theta \right) }\leq
\gamma <\infty .
\end{equation*}
\end{description}

Then, Sampling Scheme \ref{Hybrid_sscheme} satisfies the minorization
condition%
\begin{equation*}
P^{n_{0}}(v;A)\geq \gamma ^{-n_{0}p_{1}}\epsilon \nu (A),
\end{equation*}%
and for all starting values for the Markov chain%
\begin{equation*}
\left\vert \mathcal{L}\{V^{\left( n\right) }\in \cdot \}-\Pi \left\{ \cdot
\right\} \right\vert _{TV}\leq \left( 1-\delta \right) ^{\left\lfloor
n/n_{0}\right\rfloor },
\end{equation*}%
where $0<\delta <1$ and $\left\lfloor n/n_{0}\right\rfloor $ is the greatest
integer not exceeding $n/n_{0}$.
\end{theorem}

The proof is similar to Theorem 6 of \cite{mendesgen2014} and is omitted.

Sufficient conditions for the condition in Theorem \ref%
{Hybrid_sscheme_theorem_2} to be satisfied are given in Lemma 7 of \cite%
{mendesgen2014}.

\section{Examples\label{U:Examples}}

\subsection{Proposal densities for the backward MCMC steps\label{U:Proposal}}

An important issue in implementing the method is the choice of the
transition kernels (\ref{MCMCGenStates_2a}) and (\ref{MCMCGenStates_2aa}).
The user specifies proposal distributions denoted by

\begin{equation}
Q_{t}\left( x_{t},dx_{t}^{\prime }|\overline{\mathbf{x}}%
_{1:t-1},x_{t}^{-b_{t}},\tilde{x}_{t+1}^{C_{t+1}},\overline{\mathbf{a}}%
_{1:t-2},\mathbf{a}_{t-1}^{-b_{t}},\theta \right) ,
\label{eq:proposaldist_1}
\end{equation}%
for $t=1,\ldots ,T-2$ and%
\begin{equation}
Q_{T-1}\left( x_{T-1},dx_{T-1}^{\prime }|\overline{\mathbf{x}}%
_{1:T-2},x_{T-1}^{-b_{T-1}},x_{T}^{b_{T}},\overline{\mathbf{a}}_{1:T-3},%
\mathbf{a}_{T-2}^{-b_{T-1}},\theta \right) ,  \label{eq:proposaldist_1a}
\end{equation}

which are used with the target distributions (\ref{MCMCGenStates_1a}) and (%
\ref{MCMCGenStates_1aa}) to calculate the acceptance probabilities of the
Metropolis-Hastings steps in Algorithms \ref{MCMCGenStates_algorithm} and %
\ref{PG_algorithm}. Denote these acceptance probabilities by%
\begin{eqnarray}
&&\alpha _{t}\left( x_{t},x_{t}^{\prime }|\overline{\mathbf{x}}_{1:t-1},%
\mathbf{x}_{t}^{-b_{t}},\tilde{x}_{t+1}^{C_{t+1}},\overline{\mathbf{a}}%
_{1:t-2},\mathbf{a}_{t-1}^{-b_{t}},\theta \right) =1\wedge
\label{eq:accprob1} \\
&&\frac{\widehat{p}\left( dx_{t}^{\prime }|\overline{\mathbf{x}}_{1:t-1},%
\tilde{x}_{t+1}^{C_{t+1}},\overline{\mathbf{a}}_{1:t-2},\theta \right)
Q_{t}\left( x_{t}^{\prime },dx_{t}|\overline{\mathbf{x}}_{1:t-1},\mathbf{x}%
_{t}^{-b_{t}},\tilde{x}_{t+1}^{C_{t+1}},\overline{\mathbf{a}}_{1:t-2},%
\mathbf{a}_{t-1}^{-b_{t}},\theta \right) }{\widehat{p}\left( dx_{t}|%
\overline{\mathbf{x}}_{1:t-1},\tilde{x}_{t+1}^{C_{t+1}},\overline{\mathbf{a}}%
_{1:t-2},\theta \right) Q_{t}\left( x_{t},dx_{t}^{\prime }|\overline{\mathbf{%
x}}_{1:t-1},\mathbf{x}_{t}^{-b_{t}},\tilde{x}_{t+1}^{C_{t+1}},\overline{%
\mathbf{a}}_{1:t-2},\mathbf{a}_{t-1}^{-b_{t}},\theta \right) }  \notag
\end{eqnarray}%
for $t=1,\ldots ,T-2$ and%
\begin{eqnarray}
&&\alpha _{T-1}\left( x_{T-1},dx_{T-1}^{\prime }|\overline{\mathbf{x}}%
_{1:T-2},x_{T-1}^{-b_{T-1}},x_{T}^{b_{T}},\overline{\mathbf{a}}_{1:T-3},%
\mathbf{a}_{T-2}^{-b_{T-1}},\theta \right) =1\wedge  \label{eq:accprob1a} \\
&&\frac{\widehat{p}\left( dx_{T-1}^{\prime }|\overline{\mathbf{x}}_{1:T-2},%
\tilde{x}_{T}^{b_{T}},\overline{\mathbf{a}}_{1:T-3},\theta \right)
Q_{t}\left( x_{T-1}^{\prime },dx_{T-1}|\overline{\mathbf{x}}%
_{1:T-2},x_{T-1}^{-b_{T-1}},x_{T}^{b_{T}},\overline{\mathbf{a}}_{1:T-3},%
\mathbf{a}_{T-2}^{-b_{T-1}},\theta \right) }{\widehat{p}\left( dx_{T-1}|%
\overline{\mathbf{x}}_{1:T-2},\tilde{x}_{T}^{b_{T}},\overline{\mathbf{a}}%
_{1:T-3},\theta \right) Q_{t}\left( x_{T-1},dx_{T-1}^{\prime }|\overline{%
\mathbf{x}}_{1:T-2},x_{T-1}^{-b_{T-1}},x_{T}^{b_{T}},\overline{\mathbf{a}}%
_{1:T-3},\mathbf{a}_{T-2}^{-b_{T-1}},\theta \right) }.  \notag
\end{eqnarray}

This section illustrates some possible choices. To simplify notation
dependence on the parameter $\theta $ will be omitted and the conditioning
states $\tilde{x}_{t+1}^{C_{t+1}}$ for $t=1,\ldots ,T-2$ and $x_{T}^{b_{T}}$
will be denoted by $\tilde{x}_{t+1}$ for $t=1,\ldots ,T-1$. We first
consider estimates of the mean and variance of the smoothing density $%
p(x_{t}|y_{1:t},\tilde{x}_{t+1})$.

\textbf{Backward weights:} The Markov transition kernel is conditional on
the particles $\mathbf{x}_{t}^{-b_{t}}$, sampled in the filtering step.
These particles provide an approximation

\begin{equation}
\widehat{p}(x|y_{1:t},\tilde{x}_{t+1})=\frac{\sum_{i\neq
b_{t}}w_{t}^{i}f_{t}(\tilde{x}_{t+1}|x_{t}^{i})\delta (x-x_{t}^{i})}{%
\sum_{i\neq b_{t}}w_{t}^{i}f_{t}(\tilde{x}_{t+1}|x_{t}^{i})}.
\label{eq:approxsmoothingdist1}
\end{equation}

for $t=1,\ldots ,T-1$ to the smoothing density $p(x_{t}|y_{1:t},\tilde{x}%
_{t+1})$ and can be used for constructing proposal densities. This method
works well if the particles $\mathbf{x}_{t}^{-b_{t}}$ provide a good
approximation to the smoothing density. The estimates for the smoothing mean 
$\widehat{x}_{t|T}$ and smoothing covariance matrix $\widehat{S}_{t|T}$ are 
\begin{equation*}
\widehat{x}_{t|T}=\frac{\sum_{i\neq b_{t}}w_{t}^{i}f_{t}(\tilde{x}%
_{t+1}|x_{t}^{i})x_{t}^{i}}{\sum_{i\neq b_{t}}w_{t}^{i}f_{t}(\tilde{x}%
_{t+1}|x_{t}^{i})}\quad \mbox{and}\quad \widehat{S}_{t|T}=\frac{\sum_{i\neq
b_{t}}w_{t}^{i}f_{t}(\tilde{x}_{t+1}|x_{t}^{i})(x_{t}^{i}-\widehat{x}%
_{t|T})(x_{t}^{i}-\widehat{x}_{t|T})^{\prime }}{\sum_{i\neq
b_{t}}w_{t}^{i}f_{t}(\tilde{x}_{t+1}|x_{t}^{i})}
\end{equation*}

\textbf{Linearization:} This constructs a Gaussian linear approximation to
the state evolution density. Write the approximate state evolution equation
as 
\begin{equation*}
x_{t+1}=h_{t+1}+H_{t+1}x_{t}+u_{t+1},
\end{equation*}%
where $u_{t}\sim N(0,\Sigma _{t+1})$. Estimate the filtered mean, $\widehat{x%
}_{t|t}$, and covariance matrix, $\widehat{S}_{t|t}$, using the particles $%
x_{t}^{-b_{t}}$ and the forward weights $w_{t}^{-b_{t}}$: 
\begin{equation*}
\widehat{x}_{t|t}=\frac{\sum_{i\neq b_{t}}w_{t}^{i}x_{t}^{i}}{\sum_{i\neq
b_{t}}w_{t}^{i}}\quad \mbox{and}\quad \widehat{S}_{t|t}=\frac{\sum_{i\neq
b_{t}}w_{t}^{i}(x_{t}^{i}-\widehat{x}_{t|t})(x_{t}^{i}-\widehat{x}%
_{t|t})^{\prime }}{\sum_{i\neq b_{t}}w_{t}^{i}}
\end{equation*}%
The mean $x_{t|T}$ and variance $S_{t|T}$ for $t=1,\ldots ,T-1$ of the
proposals for (\ref{eq:proposaldist_1}) and (\ref{eq:proposaldist_1a}) are 
\begin{eqnarray}
x_{t|T} &=&\widehat{x}_{t|t}+\widehat{S}_{t|t}H_{t+1}^{\prime
}R_{t+1}^{-1}e_{t+1}  \label{eq:filtersmoothermean} \\
S_{t|T} &=&\widehat{S}_{t|t}-\widehat{S}_{t|t}H_{t+1}^{\prime
}R_{t+1}^{-1}H_{t+1}\widehat{S}_{t|t},  \label{eq:filtersmoothervar}
\end{eqnarray}%
where $R_{t+1}=H_{t+t}S_{t|t}H_{t+1}^{\prime }+\Sigma _{t+1}$ and $e_{t+1}=%
\tilde{x}_{t+1}-h_{t+1}-H_{t+1}\widehat{x}_{t|t}$. A similar algorithm may
be used if the state evolution equation is approximated by a Gaussian
mixture.

Although less flexible, this approach is preferred if the state evolution
density is linear and Gaussian or can be approximated arbitrarily well by a
mixture Gaussian density. An advantage of this method over the first one is
that it does not require calculating the backward weights and, therefore,
can be applied to state space models in which (\ref{eq:approxsmoothingdist1}%
) does not provide a good approximation to $p(x_{t}|y_{1:t},\tilde{x}_{t+1})$%
.

\textbf{Random walk proposal:} The two previous approaches to estimating the
variance of the smoothing density may be used to construct a random walk
proposal. In the random walk proposal we use the estimate of variance and
multiply it by a factor of $2.38^{d_{x}}/d_{x}$, where $d_{x}$ is the
dimension of the state vector $x_{t}$.

\textbf{Independent elliptical proposal:} For the independent elliptical
proposal, traditional choices of densities are a non-central Student $t$
distribution or a Gaussian distribution, where the scale and mean are
calculated using the previous methods. The computational cost for
constructing these proposal densities increases linearly with the number of
particles.

\textbf{Bootstrap proposal:} The third alternative is similar in spirit to
the bootstrap filter. The most expensive part of the MCMC moves is
evaluating $\sum_{i}w_{t-1}^{i}f(x|x_{t-1}^{i})$, present in the acceptance
probabilities (\ref{eq:accprob1})\ and (\ref{eq:accprob1a}). We suggest
using the proposal density 
\begin{equation}
q(x|\mathbf{x}_{t-1})=\sum_{i=1}^{N_{t-1}}W_{t-1}^{i}f_{t}(x|x_{t-1}^{i}).
\label{eq:propbootstrap}
\end{equation}

for $t=1,\ldots ,T-1$. For two distinct values $x$ and $x^{\prime }$, the
Metropolis-Hastings ratio is 
\begin{equation*}
1\wedge \frac{f_{t}(\tilde{x}_{t+1}|x)\,g_{t}(y_{t}|x)}{f_{t}(\tilde{x}%
_{t+1}|x^{\prime })\,g_{t}(y_{t}|x^{\prime })},
\end{equation*}%
as the proposal density (\ref{eq:propbootstrap}) and the sum in the target
densities (\ref{MCMCGenStates_1a}) and (\ref{MCMCGenStates_1aa}) cancel out.

This method is faster than the previous ones, but it does not take into
account $\tilde{x}_{t+1}$ or $y_{t}$ to construct the proposal. Furthermore,
the only assumption about the state evolution equation is that it can be
sampled from and evaluated up to a normalizing constant.

\subsection{Nonlinear state space model\label{U:nonlinear_ssm}}

The goal of this example is to evaluate the performance of the algorithm for
several combinations of the number of particles $N$ and MCMC iterations $C$.
We consider the following nonlinear state-space model used by many authors
including \cite{gordonetal1993}, \cite{kitagawa1996}, and \cite%
{andrieuetal2010}: 
\begin{eqnarray*}
y_{t} &=&\frac{x_{t}^{2}}{20}+\sigma \varepsilon _{t} \\
x_{t+1} &=&\frac{x_{t}}{2}+25\frac{x_{t}}{1+x_{t}^{2}}+8\cos (1.2\,t)+\tau
\eta _{t+1},
\end{eqnarray*}%
where $\varepsilon _{t}$ and $\eta _{t}$ are standard Gaussian random
variables and $x_{1}\sim N(0,5)$. We choose an inverse Gamma prior for both $%
\sigma ^{2}$ and $\tau ^{2}$ with shape $1$ and scale $0.1$.

We simulate 50 observations with parameters $\sigma ^{2}=1$ and $\tau
^{2}=10 $. The bootstrap filter samples the particles from the state
evolution equation, while the bootstrap MCMC proposal samples the states
from (\ref{eq:propbootstrap}). The variance term in the random walk proposal
is calculated using (\ref{eq:approxsmoothingdist1}) and scaled by the factor 
$2.38$. Despite being sub-optimal, these choices are very general and only
require that one can sample from the state density and can evaluate both the
observation and state densities. We avoid using a Gaussian independent
proposal as it provides a poor approximation to the bimodal target.

\begin{figure}[ht]
\centering
\begin{subfigure}[b]{0.5\textwidth}
            \includegraphics[width=\textwidth]{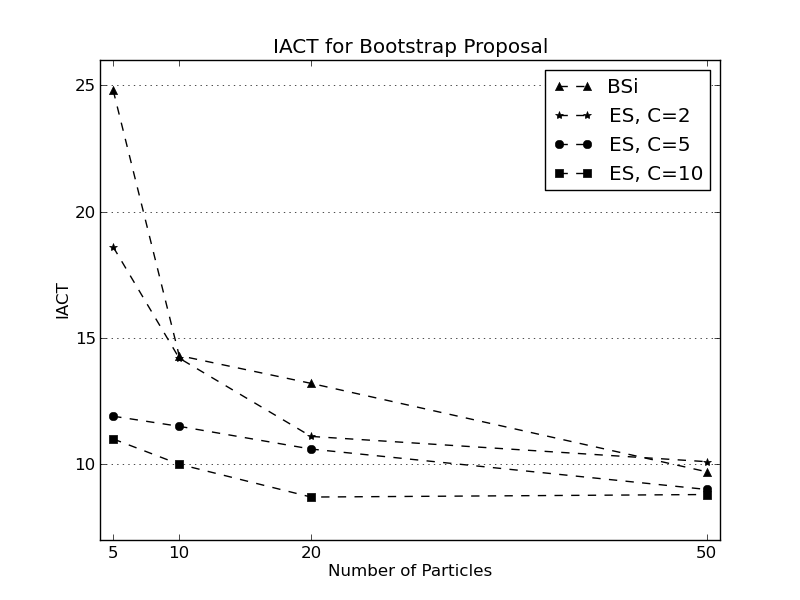}
            \caption{}
            \label{fig:nonlinboot}
    \end{subfigure} ~ 
\begin{subfigure}[b]{0.5\textwidth}
            \includegraphics[width=\textwidth]{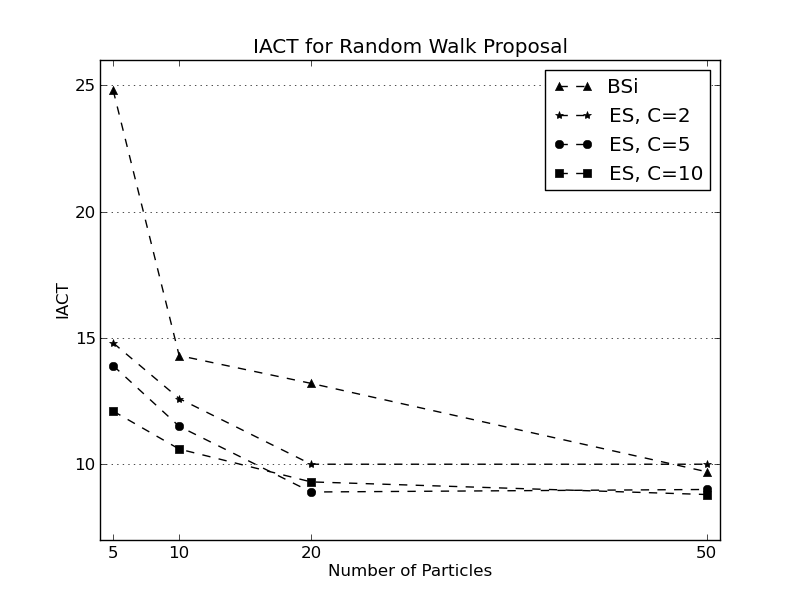}
            \caption{}
            \label{fig:nonlinrw}
    \end{subfigure}
\caption{Nonlinear state space model. Maximum of the IACT of $\protect\sigma$
and $\protect\tau$ for different choices of number of particles
(N=5,10,20,50) and MCMC steps (C=2,5,10) for the extended space particle
Gibbs sampler (ES). Also shown is the performance of the backward simulation
approach in \protect\cite{lindstenschon2012} (BSi). Panels (a) and (b) show
the result for the ``bootstrap'' proposal and the random walk proposal with
backward weights, respectively.}
\end{figure}

We generate 100,000 iterations and discard the initial 10,000 as warm up.
The performance of each method is measured as the maximum IACT of the $%
\sigma $ and $\tau $ iterates, i.e., $\max (IACT(\tau ),IACT(\sigma ))$
where $IACT(\theta )$ is the integrated autocorrelation time estimate of a
parameter $\theta $. In the simulations, we take $N=5,10,20,50$ and $%
C=2,5,10 $. Figures \ref{fig:nonlinboot} and \ref{fig:nonlinrw} show the
results for the particle Gibbs sampler using a bootstrap proposal and a
random walk proposal, respectively. Both results are compared with the
particle Gibbs with backward simulation method proposed by \cite%
{lindstenschon2012}. To distinguish between the methods we will refer to our
approach as the extended space particle Gibbs sampler. The IACTs are
calculated using overlapping batch means \citep[see,
e.g.][]{jonesetal2006} using 90,000 samples and block size 300.

The performance of the samplers improves as the number of particles and
number of MCMC moves increase. The bootstrap proposal performs worse than
the random walk proposal for a fixed number of particles and MCMC steps. In
practice, the difference between the IACTs is negligible as the number of
particles increases. In this simulation study, the random walk proposal is
between 2.3 and 6.7 times slower than the bootstrap proposal, depending on
the number of particles and MCMC moves.

\subsection{Stochastic Volatility Model\label{U:SV}}

In this example we fit a stochastic volatility (SV) model for the
Pound/Dollar daily exchange rates from 1-Oct-1981 to 26-Jun-1985 %
\citep[see][ Sec. 14.4]{durbinkoopman}. The observation and state transition
equations are 
\begin{equation*}
y_{t}=e^{h_{t}/2}\varepsilon _{t}\quad \mbox{and}\quad h_{t+1}=\mu +\phi
(h_{t}-\mu )+\tau \eta _{t+1},
\end{equation*}%
with $(\varepsilon _{t},\eta _{t})$ independently distributed standard
Gaussian random variables. In this model, $h_{t}$ is the log-volatility at
time $t$, $\mu $ is the mean of the log-volatility, $\tau $ its standard
deviation and $\phi $ the autoregressive parameter. The specification of the
SV model is completed by assuming the distribution of the initial state $%
h_{1}\sim N\left( \mu ,\tau ^{2}/(1-\phi ^{2})\right) $. We are interested
in performing Bayesian inference for the parameters of this model. To
complete the Bayesian specification, we use the following priors. The
autocorrelation parameter $\phi $ is uniform on $(-1,1)$, the state standard
deviation $\tau $ has a half-$t$ distribution with degrees of freedom $d=4$ %
\citep{gelman2006}, and the log-volatility mean $\mu \sim N(0,2^{2})$.

We reparametrize the model as $y_{t}=\exp \{\mu +\tau x_{t}\}\varepsilon
_{t} $ where $x_{t}=(h_{t}-\mu )/\tau $ and $x_{t+1}=\phi x_{t}+\eta _{t+1}$
with $x_{1}\sim N(0,1/(1-\phi ^{2}))$, using independent Metropolis steps
within the Gibbs sampler to draw from $p(\mu ,\tau |x_{1:T},y_{1:T})$ and $%
p(\phi |x_{1:T})$. The independent proposals are calculated using the
Laplace approximation of the conditional densities and yield an acceptance
rate close to 90\%.

We consider an adapted particle filter using the \textit{optimal} importance
densities described in \cite{doucetetal2000} and five different proposal
densities for the MCMC step: a random walk with variance calculated using (%
\ref{eq:approxsmoothingdist1}), a random walk with variance calculated using
(\ref{eq:filtersmoothervar}), a Gaussian independent proposal with mean and
variances calculated using (\ref{eq:approxsmoothingdist1}), a Gaussian
independent proposal with mean and variance calculated using (\ref%
{eq:filtersmoothermean}) and (\ref{eq:filtersmoothervar}), and the
independent proposal (\ref{eq:propbootstrap}). We compare the performance of
the proposals for several different combinations of the number of particles $%
N$ and MCMC iterations $C$.

We run the extended space particle Gibbs sampler for 50,000 iterations using
the first 5,000 as warmup and calculate the IACT (inefficiency factor) using
the overlapping batch means method with a bandwidth of 213 samples. In all
the simulations, the posterior means and variances are consistent with
values previously found in the literature 
\citep[see, e.g.][ Sec
14.4]{durbinkoopman}. The posterior means for $\mu $, $\tau $ and $\phi $
are respectively $-.952$, $.180$ and $.971$, while the posterior standard
deviations are $.1997$, $.0351$ and $.0126$, respectively. Table \ref%
{t:pound-dollar} shows the IACT and relative time to run each algorithm for
each combination of $N$ and $C$. We display the relative time to run each of
the algorithms compared to the bootstrap MCMC proposal. The bootstrap
proposal is used as the baseline for time because it is fastest; it avoids $%
C\times N\times T$ state evolution density evaluations when compared to the
independent sampler or the random walk sampler.

Table \ref{t:pound-dollar} shows that in this example all the algorithms
perform similarly, given the number of particles and MCMC steps. As the
number of particles increases the efficiencies of the methods decrease
steadily. Increasing the number of MCMC steps, however, does not have a
significant impact on the IACT. One possible explanation is that only a few
iterations are enough to break the dependence structure of the conditional
sequential Monte Carlo and increasing the number of steps is irrelevant.
Finally, a bootstrap MCMC proposal with $N$=400 and $C$=5 yields IACT($\mu $%
) = 51, IACT($\tau $) = 29 and IACT($\phi $) = 23, showing that even
increasing the number of particles eight-fold does not decrease the
inefficiencies much.

\begin{table}[tbp]
\centering
{\footnotesize \ 
\begin{tabular}{cclcccc}
\hline\hline
$N$ & $C$ & Method & IACT($\mu$) & IACT($\phi$) & IACT($\tau$) & Relative
Time \\ \hline
\multirow{15}{*}{10} & \multirow{5}{*}{5} & Boot & 132.9 & 82.2 & 66.9 & 1.00
\\ 
&  & RW1 & 122.3 & 82.0 & 63.3 & 2.81 \\ 
&  & RW2 & 116.6 & 77.3 & 61.4 & 2.36 \\ 
&  & Ind1 & 141.3 & 85.1 & 72.1 & 3.10 \\ 
&  & Ind2 & 111.7 & 67.3 & 55.5 & 2.55 \\ \cline{2-7}
& \multirow{5}{*}{10} & Boot & 139.3 & 74.7 & 65.8 & 1.00 \\ 
&  & RW1 & 108.3 & 71.3 & 54.4 & 3.29 \\ 
&  & RW2 & 140.3 & 75.7 & 64.2 & 3.06 \\ 
&  & Ind1 & 122.4 & 77.9 & 60.7 & 3.65 \\ 
&  & Ind2 & 119.2 & 82.0 & 66.1 & 3.20 \\ \cline{2-7}
& \multirow{5}{*}{20} & Boot & 105.1 & 75.2 & 58.5 & 1.00 \\ 
&  & RW1 & 125.7 & 80.1 & 63.5 & 4.84 \\ 
&  & RW2 & 116.8 & 71.9 & 56.8 & 4.68 \\ 
&  & Ind1 & 121.8 & 80.7 & 63.2 & 5.44 \\ 
&  & Ind2 & 116.5 & 69.4 & 58.5 & 5.22 \\ \hline
\multirow{15}{*}{20} & \multirow{5}{*}{5} & Boot & 102.3 & 63.3 & 52.9 & 1.00
\\ 
&  & RW1 & 113.2 & 56.9 & 47.9 & 3.36 \\ 
&  & RW2 & 98.0 & 50.9 & 44.8 & 2.74 \\ 
&  & Ind1 & 90.3 & 56.4 & 44.8 & 3.70 \\ 
&  & Ind2 & 82.4 & 50.3 & 39.2 & 2.99 \\ \cline{2-7}
& \multirow{5}{*}{10} & Boot & 91.8 & 60.7 & 49.5 & 1.00 \\ 
&  & RW1 & 116.5 & 62.9 & 55.9 & 4.24 \\ 
&  & RW2 & 96.1 & 54.3 & 45.5 & 3.81 \\ 
&  & Ind1 & 111.8 & 65.1 & 53.7 & 4.62 \\ 
&  & Ind2 & 100.9 & 51.1 & 49.5 & 4.10 \\ \cline{2-7}
& \multirow{5}{*}{20} & Boot & 95.79 & 56.0 & 44.48 & 1.00 \\ 
&  & RW1 & 115.9 & 58.1 & 49.5 & 6.81 \\ 
&  & RW2 & 116.7 & 58.0 & 48.4 & 6.59 \\ 
&  & Ind1 & 89.8 & 54.7 & 44.3 & 7.36 \\ 
&  & Ind2 & 90.4 & 53.4 & 43.2 & 6.89 \\ \hline
\multirow{15}{*}{50} & \multirow{5}{*}{5} & Boot & 74.9 & 46.5 & 37.5 & 1.00
\\ 
&  & RW1 & 73.2 & 43.1 & 36.2 & 4.62 \\ 
&  & RW2 & 93.6 & 48.1 & 41.6 & 3.25 \\ 
&  & Ind1 & 73.1 & 41.9 & 33.7 & 5.02 \\ 
&  & Ind2 & 59.9 & 45.0 & 35.3 & 3.39 \\ \cline{2-7}
& \multirow{5}{*}{10} & Boot & 78.6 & 57.2 & 41.4 & 1.00 \\ 
&  & RW1 & 77.5 & 43.4 & 35.3 & 5.21 \\ 
&  & RW2 & 68.7 & 41.4 & 32.4 & 4.74 \\ 
&  & Ind1 & 97.3 & 52.0 & 45.9 & 6.04 \\ 
&  & Ind2 & 86.2 & 39.0 & 34.7 & 5.13 \\ \cline{2-7}
& \multirow{5}{*}{20} & Boot & 72.2 & 48.9 & 38.3 & 1.00 \\ 
&  & RW1 & 73.2 & 43.0 & 34.5 & 9.32 \\ 
&  & RW2 & 94.8 & 44.8 & 38.6 & 8.48 \\ 
&  & Ind1 & 76.7 & 44.9 & 37.0 & 9.47 \\ 
&  & Ind2 & 70.0 & 43.5 & 36.2 & 8.84 \\ \hline\hline
\end{tabular}%
}
\caption{Stochastic volatiliy model. Inefficiency factors (IACT) for the
extended space particle Gibbs sampler, using five different proposals for
the MCMC for the states and different number of particles $N$ and number of
MCMC moves $C$. The MCMC proposals are random walk with scale calculated
using the particles (RW1), random walk with scale calculated using
linearization (RW2), Gaussian proposal with mean and variance calculated
using the particles (Ind1), Gaussian proposal with mean and variance
calculated using linearization (Ind2), and the bootstrap proposal (Boot).
Relative Time shows the relative time to run each proposal when compared to
the bootstrap (Boot), given the number of particles and MCMC moves.}
\label{t:pound-dollar}
\end{table}

\subsection{Binomial regression model with time-varying coefficients\label%
{U:Binomialreg}}

Consider the state space model with binomial observations 
\begin{equation}
Y_{t}|\boldsymbol{x}_{t}\sim \mbox{Binomial}(n_{t},p_{t}),\quad \log \frac{%
p_{t}}{1-p_{t}}=\beta _{0}+\beta _{1,t}z_{1,t}+\cdots +\beta _{m,t}z_{m,t}
\label{eq:binregmodel}
\end{equation}%
where $\boldsymbol{z}_{t}=(z_{1,t},\dots ,z_{m,t})^{\prime }$ is vector of
covariates. We take the intercept $\beta _{0}$ to be fixed for all $t$ with
a prior $N\left( 0,5^{2}\right) $, but allow the coefficients $\boldsymbol{%
\beta }_{t}=(\beta _{1,t},\dots ,\beta _{m,t})^{\prime }$ to evolve over
time by using the random walk prior $\boldsymbol{\beta }_{t}=\boldsymbol{%
\beta }_{t-1}+\boldsymbol{\tau \eta }_{t}$ for $t=1,\ldots ,T,$with $%
\boldsymbol{\eta }_{t}\sim N\left( 0,I_{m}\right) $ and $\boldsymbol{\tau }=%
\mbox{diag}(\tau _{1},\dots ,\tau _{m})$. We use the prior $\tau
_{i}^{2}\sim IG(1,.5)$ for $i=1,..,m$, with the $\tau _{i}$ independent
apriori. We also generate independent values of $n_{t}\sim \mbox{Binomial}%
(100,0.5)$ for $t=1,\ldots ,T$, which gave values of $n_{t}$ lying in the
interval $\left( 35,65\right) $. We generate independent values of the
covariates $z_{i,t}\sim U\left( -1,1\right) $ for $i=1,\ldots ,m$ and $%
t=1,\ldots ,T$. We generate $T=200$ observations from model (\ref%
{eq:binregmodel}), setting $\beta _{0}=0.5$ and $\tau _{i}=0.6$ and $\beta
_{i,0}=0$ for $i=1,\ldots ,m$, and we take the number of covariates $m=4$.
The minimum generated values of $p_{t}$ were close to zero and the maximum
values were close to one.

The extended space particle Gibbs specification uses a bootstrap particle
filter with the bootstrap\ MCMC proposal for the states. We vary the number
of particles and MCMC moves and estimate the largest IACT among the $\tau
_{i}$s ($\max_{i=1,\dots ,p}IACT(\tau _{i})$). We run 100,000 iterations of
the Gibbs sampler and discard the initial 5,000 as warm up. The remaining
95,000 samples are used to calculate the IACT using the overlapping batch
means method, using a window size of 309 samples.

\begin{figure}[ht]
\centering
\begin{subfigure}[b]{0.5\textwidth}
            \includegraphics[width=\textwidth]{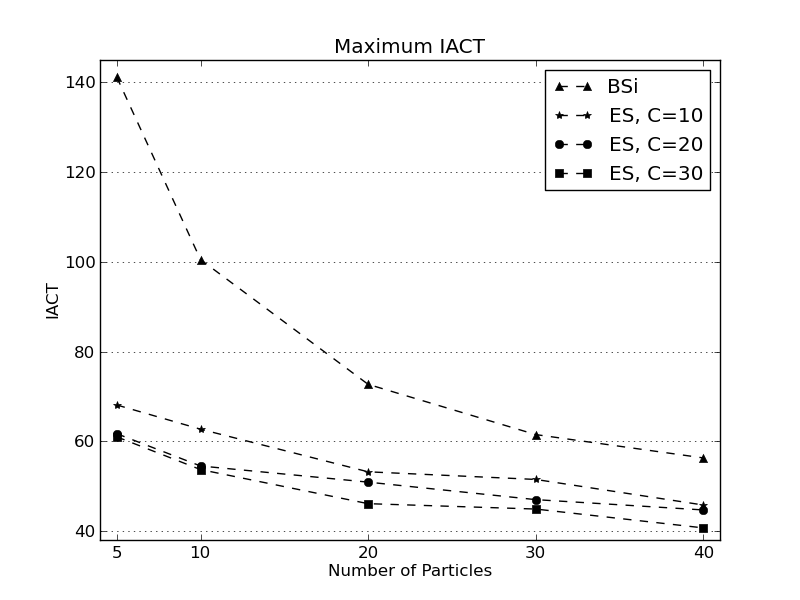}
            \caption{}
            \label{fig:binompartial}
    \end{subfigure} ~ 
\begin{subfigure}[b]{0.5\textwidth}
            \includegraphics[width=\textwidth]{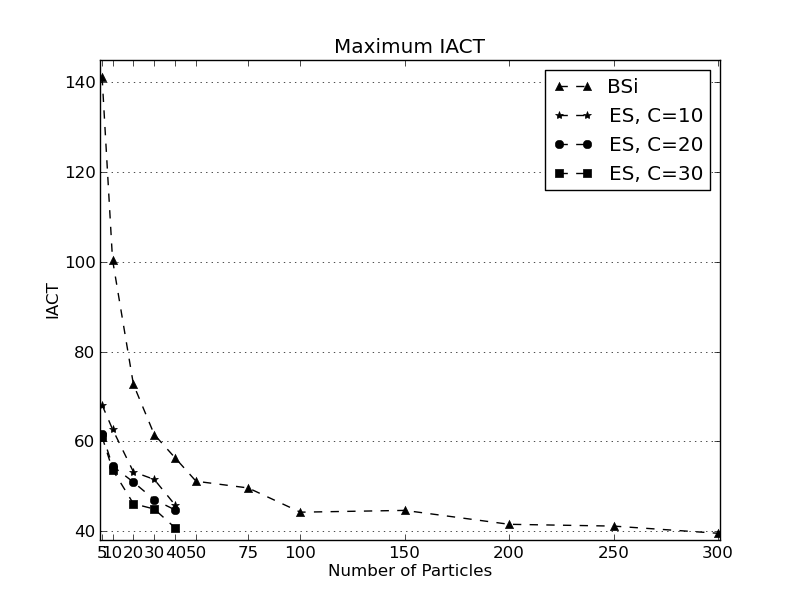}
            \caption{}
            \label{fig:binomfull}
    \end{subfigure}
\caption{Binomial regression model. Maximum IACT among the $\protect\tau_i$s
($i=1,2,3,4$) for different choices of number of particles (N=5,10,20,40)
and MCMC steps (C=10,20,30) for the extended space particle Gibbs sampler
(ES). Also shown is the performance of the backward simulation approach in 
\protect\cite{lindstenschon2012} (BSi). We also consider $%
N=50,75,100,150,200,250$ for the backward simulation approach (BSi) in panel
(b). The results for the extended space particle Gibbs sampler is displayed
in panel (a). Panel (b) shows the results for the backward simulation
algorithm, compared with our new approach. The IACT is calculated using
overlap batch means with 95,000 samples and window size 309.}
\end{figure}

Figure \ref{fig:binompartial} displays the IACTs for our method using $%
N=5,10,20,30,40$ and $C=10,20,30$ and compares it with the backward
simulation algorithm of \cite{lindstenschon2012} using the same number of
particles. Figure \ref{fig:binomfull} shows the IACTs for the backward
simulation method for $N=5,10,20,30,40,50,75,100,150,200,250,300$ and
compares with the results using the new approach. The inefficiency factors
for the new approach converge to a minimum faster than backward simulation.
The extended support approach yields a minimal IACT with $N=40$ and $C=30$,
while the backward simulation takes around $N=250$ particles to achieve this
value and twice the time in our particular, general, specification using the
Julia programming language.

\bibliographystyle{abbrvnat}
\bibliography{pfmcmc}

\appendix{}

\section{Proofs of lemmas}

\textit{Proof of Lemma \ref{MHRatio_lemma}.}

To prove Part (i), integrate $\Pi \left( d\mathbf{u}_{1:T}|\theta \right) $
over $\mathbf{x}_{T}^{-b_{T}}$ and sum over $\overline{\mathbf{a}}_{T-1}$ to
get%
\begin{eqnarray}
&&\Pi \left( d\overline{\mathbf{x}}_{1:T-1},dx_{T}^{b_{T}},\overline{\mathbf{%
a}}_{1:T-2},\mathbf{b}_{1:T},d\widetilde{\mathbf{x}}_{1:T-1}|\theta \right) 
\notag \\
&=&\pi \left( \widetilde{x}_{1}^{C_{1}},\ldots ,\widetilde{x}%
_{T-1}^{C_{T-1}},x_{T}^{b_{T}}|\theta \right) d\widetilde{x}%
_{1}^{C_{1}}\ldots d\widetilde{x}_{T-1}^{C_{T-1}}dx_{T}^{b_{T}}\left( \frac{1%
}{N_{T}}\right)  \notag \\
&&\frac{\Psi \left( d\overline{\mathbf{x}}_{1:T-1},\overline{\mathbf{a}}%
_{1:T-2}|\theta \right) }{M_{1}\left( dx_{1}^{b_{1}}|y_{1},\theta \right)
\dprod_{t=2}^{T-1}\left\{ W_{t-1}^{a_{t-1}^{b_{t}}}M_{t}\left(
dx_{t}^{b_{t}}|y_{t},x_{t-1}^{a_{t-1}^{b_{t}}},\theta \right) \right\} } 
\notag \\
&&\left\{ \dprod_{t=2}^{T-1}\frac{w_{t-1}^{a_{t-1}^{b_{t}}}f_{t}\left(
x_{t}^{b_{t}}|x_{t-1}^{a_{t-1}^{b_{t}}},\theta \right) }{%
\dsum_{i=1}^{N_{t-1}}w_{t-1}^{i}f_{t}\left( x_{t}^{b_{t}}|x_{t-1}^{i},\theta
\right) }\right\} \left( \frac{1}{N_{T-1}}\right)  \notag \\
&&\dprod_{j=2}^{C_{T-1}}K_{T-1}\left( \widetilde{x}_{T-1}^{j},d\widetilde{x}%
_{T-1}^{j-1}|\overline{\mathbf{x}}_{1:T-2},\mathbf{x}%
_{T-1}^{-b_{T-1}},x_{T}^{b_{T}},\overline{\mathbf{a}}_{1:T-3},\mathbf{a}%
_{T-2}^{-b_{T-1}},\theta \right)  \notag \\
&&K_{T-1}\left( \widetilde{x}_{T-1}^{1},dx_{T-1}^{b_{T-1}}|\overline{\mathbf{%
x}}_{1:T-2},\mathbf{x}_{T-1}^{-b_{T-1}},x_{T}^{b_{T}},\overline{\mathbf{a}}%
_{1:T-3},\mathbf{a}_{T-2}^{-b_{T-1}},\theta \right)  \notag \\
&&\dprod_{t=1}^{T-2}\left\{ \left( \frac{1}{N_{t}}\right)
\dprod_{j=2}^{C_{t}}K_{t}\left( \widetilde{x}_{t}^{j},d\widetilde{x}%
_{t}^{j-1}|\overline{\mathbf{x}}_{1:t-1},\mathbf{x}_{t}^{-b_{t}},\widetilde{x%
}_{t+1}^{Ct+1},\overline{\mathbf{a}}_{1:t-2},\mathbf{a}_{t-1}^{-b_{t}},%
\theta \right) \right.  \notag \\
&&\left. K_{t}\left( \widetilde{x}_{t}^{1},dx_{t}^{b_{t}}|\overline{\mathbf{x%
}}_{1:t-1},\mathbf{x}_{t}^{-b_{t}},\widetilde{x}_{t+1}^{Ct+1},\overline{%
\mathbf{a}}_{1:t-2},\mathbf{a}_{t-1}^{-b_{t}},\theta \right) \right\} .
\label{MHRatio_lemma_0a}
\end{eqnarray}%
Now sum over $a_{T-2}^{b_{T-1}}$, integrate over $x_{T-1}^{b_{T-1}},%
\widetilde{x}_{T-1}^{1},\ldots ,\widetilde{x}_{T-1}^{C_{T-1}-1}$, integrate
over $\mathbf{x}_{T-1}^{-b_{T-1}}$, sum over $\mathbf{a}_{T-2}^{-b_{T-1}}$,
and sum over $b_{T-1}$ to get%
\begin{eqnarray}
&&\Pi \left( d\overline{\mathbf{x}}_{1:T-2},dx_{T}^{b_{T}},\overline{\mathbf{%
a}}_{1:T-3},b_{1:T-2},b_{T},d\widetilde{\mathbf{x}}_{1:T-2},d\widetilde{x}%
_{T-1}^{C_{T-1}}|\theta \right)  \notag \\
&=&\pi \left( \widetilde{x}_{1}^{C_{1}},\ldots ,\widetilde{x}%
_{T-1}^{C_{T-1}},x_{T}^{b_{T}}|\theta \right) d\widetilde{x}%
_{1}^{C_{1}}\ldots d\widetilde{x}_{T-1}^{C_{T-1}}dx_{T}^{b_{T}}\left( \frac{1%
}{N_{T}}\right)  \notag \\
&&\frac{\psi \left( \overline{\mathbf{x}}_{1},\ldots ,\overline{\mathbf{x}}%
_{T-2},\overline{\mathbf{a}}_{1},\ldots ,\overline{\mathbf{a}}_{T-3}|\theta
\right) d\overline{\mathbf{x}}_{1}\ldots d\overline{\mathbf{x}}_{T-2}}{%
M_{1}\left( x_{1}^{b_{1}}|y_{1},\theta \right) \dprod_{t=2}^{T-2}\left\{
w_{t-1}^{a_{t-1}^{b_{t}},\ast }M_{t}\left(
x_{t}^{b_{t}}|y_{t},x_{t-1}^{a_{t-1}^{b_{t}}}\right) \right\} }  \notag \\
&&\dprod_{t=2}^{T-2}\frac{w_{t-1}^{a_{t-1}^{b_{t}}}f_{t}\left(
x_{t}^{b_{t}}|x_{t-1}^{a_{t-1}^{b_{t}}},\theta \right) }{%
\dsum_{i=1}^{N_{t-1}}w_{t-1}^{i}f_{t}\left( x_{t}^{b_{t}}|x_{t-1}^{i},\theta
\right) }  \notag \\
&&\dprod_{t=1}^{T-2}\left\{ \left( \frac{1}{N_{t}}\right)
\dprod_{j=2}^{C_{t}}K_{t}\left( \widetilde{x}_{t}^{j},d\widetilde{x}%
_{t}^{j-1}|\overline{\mathbf{x}}_{1},\ldots ,\overline{\mathbf{x}}_{t-1},%
\mathbf{x}_{t}^{-b_{t}},\widetilde{x}_{t+1}^{Ct+1},\overline{\mathbf{a}}%
_{1},\ldots ,\overline{\mathbf{a}}_{t-2},\mathbf{a}_{t-1}^{-b_{t}},\theta
\right) \right.  \notag \\
&&\left. K_{t}\left( \widetilde{x}_{t}^{1},dx_{t}^{b_{t}}|\overline{\mathbf{x%
}}_{1},\ldots ,\overline{\mathbf{x}}_{t-1},\mathbf{x}_{t}^{-b_{t}},%
\widetilde{x}_{t+1}^{Ct+1},\overline{\mathbf{a}}_{1},\ldots ,\overline{%
\mathbf{a}}_{t-1},\theta \right) \right\} .  \label{MHRatio_lemma_0a1}
\end{eqnarray}%
The expression in (\ref{MHRatio_lemma_0a1}) shows that we can repeatedly for 
$t=T-2,\ldots ,2$ do the following: sum over $a_{t-1}^{b_{t}}$, integrate
over $x_{t}^{b_{t}},\widetilde{x}_{t}^{1},\ldots ,\widetilde{x}%
_{t}^{C_{t}-1} $, integrate over $\mathbf{x}_{t}^{-b_{t}}$, sum over $%
\mathbf{a}_{t-1}^{-b_{t}}$, and sum over $b_{t}$ to get%
\begin{eqnarray}
&&\Pi \left( d\overline{\mathbf{x}}_{1:t-1},dx_{T}^{b_{T}},\overline{\mathbf{%
a}}_{1:t-2},\mathbf{b}_{1:t-1},b_{T},d\widetilde{\mathbf{x}}_{1:t-1},d%
\widetilde{x}_{t}^{C_{t}},\ldots ,d\widetilde{x}_{T-1}^{C_{T-1}}|\theta
\right)  \notag \\
&=&\pi \left( \widetilde{x}_{1}^{C_{1}},\ldots ,\widetilde{x}%
_{T-1}^{C_{T-1}},x_{T}^{b_{T}}|\theta \right) d\widetilde{x}%
_{1}^{C_{1}}\ldots d\widetilde{x}_{T-1}^{C_{T-1}}dx_{T}^{b_{T}}\left( \frac{1%
}{N_{T}}\right)  \notag \\
&&\frac{\Psi \left( d\overline{\mathbf{x}}_{1:t-1},\overline{\mathbf{a}}%
_{1:t-2}|\theta \right) }{M_{1}\left( dx_{1}^{b_{1}}|y_{1},\theta \right)
\dprod_{s=2}^{t-1}\left\{ W_{s-1}^{a_{s-1}^{b_{t}}}M_{s}\left(
dx_{s}^{b_{s}}|y_{s},x_{s-1}^{a_{s-1}^{b_{s}}}\right) \right\} }  \notag \\
&&\dprod_{s=2}^{t-1}\frac{w_{s-1}^{a_{s-1}^{b_{s}}}f_{t}\left(
x_{s}^{b_{s}}|x_{s-1}^{a_{s-1}^{b_{s}}},\theta \right) }{%
\dsum_{i=1}^{N_{s-1}}w_{s-1}^{i}f_{s}\left( x_{s}^{bs}|x_{s-1}^{i},\theta
\right) }  \notag \\
&&\dprod_{s=1}^{t-1}\left\{ \left( \frac{1}{N_{s}}\right)
\dprod_{j=2}^{C_{s}}K_{s}\left( \widetilde{x}_{s}^{j},d\widetilde{x}%
_{s}^{j-1}|\overline{\mathbf{x}}_{1:t-1},\mathbf{x}_{s}^{-b_{s}},\widetilde{x%
}_{s+1}^{Cs+1},\overline{\mathbf{a}}_{1:t-2},\mathbf{a}_{s-1}^{-b_{s}},%
\theta \right) \right.  \notag \\
&&\left. K_{s}\left( \widetilde{x}_{s}^{1},dx_{s}^{b_{s}}|\overline{\mathbf{x%
}}_{1:t-1},\mathbf{x}_{s}^{-b_{s}},\widetilde{x}_{s+1}^{Cs+1},\overline{%
\mathbf{a}}_{1:t-2},\mathbf{a}_{s-1}^{-b_{s}},\theta \right) \right\} .
\label{MHRatio_lemma_0b}
\end{eqnarray}%
For $t=1$ this simplifies to 
\begin{eqnarray}
&&\Pi \left( d\overline{\mathbf{x}}_{1},dx_{T}^{b_{T}},b_{1},b_{T},d%
\widetilde{\mathbf{x}}_{1},d\widetilde{x}_{2}^{C_{2}},\ldots ,d\widetilde{x}%
_{T-1}^{C_{T-1}}|\theta \right)  \notag \\
&=&\pi \left( \widetilde{x}_{1}^{C_{1}},\ldots ,\widetilde{x}%
_{T-1}^{C_{T-1}},x_{T}^{b_{T}}|\theta \right) d\widetilde{x}%
_{1}^{C_{1}}\ldots d\widetilde{x}_{T-1}^{C_{T-1}}dx_{T}^{b_{T}}\left( \frac{1%
}{N_{T}}\right)  \label{MHRatio_lemma_0c} \\
&&\frac{\psi \left( \overline{\mathbf{x}}_{1}|\theta \right) d\overline{%
\mathbf{x}}_{1}}{M_{1}\left( x_{1}^{b_{1}}|y_{1},\theta \right) }\frac{1}{%
N_{1}}\dprod_{j=2}^{C_{t}}K_{1}\left( \widetilde{x}_{1}^{j},d\widetilde{x}%
_{1}^{j-1}|\mathbf{x}_{1}^{-b_{t}},\widetilde{x}_{2}^{C_{2}},\theta \right)
K_{1}\left( \widetilde{x}_{1}^{1},dx_{1}^{b_{1}}|\mathbf{x}_{1}^{-b_{1}},%
\widetilde{x}_{t+1}^{Ct+1},\theta \right) .  \notag
\end{eqnarray}%
Now integrate over $x_{1}^{b_{1}},\widetilde{x}_{1}^{1},\ldots ,\widetilde{x}%
_{1}^{C_{t1}-1}$, integrate over $\mathbf{x}_{1}^{-b_{1}},$ and sum over $%
b_{1}$ to get%
\begin{eqnarray*}
&&\Pi \left( dx_{T}^{b_{T}},b_{T},d\widetilde{\mathbf{x}}_{1}^{C_{1}},d%
\widetilde{x}_{2}^{C_{2}},\ldots ,d\widetilde{x}_{T-1}^{C_{T-1}}|\theta
\right) \\
&=&\pi \left( \widetilde{x}_{1}^{C_{1}},\ldots ,\widetilde{x}%
_{T-1}^{C_{T-1}},x_{T}^{b_{T}}|\theta \right) d\widetilde{x}%
_{1}^{C_{1}}\ldots d\widetilde{x}_{T-1}^{C_{T-1}}dx_{T}^{b_{T}}\left( \frac{1%
}{N_{T}}\right) ,
\end{eqnarray*}%
as required.

For Part (ii), we first note that from Assumption \ref{assumption:support},
the probability measure $\Pi \left( d\mathbf{u}_{1:T}|\theta \right) $ is
equivalent to the probability measure%
\begin{eqnarray*}
&&\Pi ^{\ast }\left( d\mathbf{u}_{1:T}|\theta \right) \mathrel{:=} \\
&&\pi \left( d\widetilde{x}_{1}^{C_{1}}|\widetilde{x}_{2}^{C_{2}},\theta
\right) \dprod_{t=2}^{T-1}\hat{p}\left( d\widetilde{x}_{t}^{C_{t}}|\overline{%
\mathbf{x}}_{1:t-1},\tilde{x}_{t+1}^{C_{t+1}},\overline{\mathbf{a}}%
_{1:t-2},\theta \right) \pi \left( dx_{T}^{b_{T}}|\theta \right) \left( 
\frac{1}{N_{T}}\right) \\
&&\frac{\Psi \left( d\overline{\mathbf{x}}_{1:T},\overline{\mathbf{a}}%
_{1:T-1}|\theta \right) }{M_{1}\left( dx_{1}^{b_{1}}|y_{1},\theta \right)
\dprod_{t=2}^{T}\left\{ W_{t-1}^{a_{t-1}^{b_{t}}}M_{t}\left(
dx_{t}^{b_{t}}|y_{t},x_{t-1}^{a_{t-1}^{b_{t}}},\theta \right) \right\} } \\
&&\dprod_{t=2}^{T}\frac{w_{t-1}^{a_{t-1}^{b_{t}}}f_{t}\left(
x_{t}^{b_{t}}|x_{t-1}^{a_{t-1}^{b_{t}}},\theta \right) }{%
\dsum_{i=1}^{N_{t-1}}w_{t-1}^{i}f_{t}\left( x_{t}^{b_{t}}|x_{t-1}^{i},\theta
\right) } \\
&&\left( \frac{1}{N_{T-1}}\right) \dprod_{j=2}^{C_{T-1}}K_{T-1}\left( 
\widetilde{x}_{T-1}^{j},d\widetilde{x}_{T-1}^{j-1}|\overline{\mathbf{x}}%
_{1:T-2},\mathbf{x}_{T-1}^{-b_{T-1}},x_{T}^{b_{T}},\overline{\mathbf{a}}%
_{1:T-3},\mathbf{a}_{T-2}^{-b_{T-1}},\theta \right) \\
&&K_{T-1}\left( \widetilde{x}_{T-1}^{1},dx_{T-1}^{b_{T-1}}|\overline{\mathbf{%
x}}_{1:T-2},\mathbf{x}_{T-1}^{-b_{T-1}},x_{T}^{b_{T}},\overline{\mathbf{a}}%
_{1:T-3},\mathbf{a}_{T-2}^{-b_{T-1}},\theta \right) \\
&&\dprod_{t=1}^{T-2}\left\{ \left( \frac{1}{N_{t}}\right)
\dprod_{j=2}^{C_{t}}K_{t}\left( \widetilde{x}_{t}^{j},d\widetilde{x}%
_{t}^{j-1}|\overline{\mathbf{x}}_{1:t-1},\mathbf{x}_{t}^{-b_{t}},\widetilde{x%
}_{t+1}^{Ct+1},\overline{\mathbf{a}}_{1:t-2},\mathbf{a}_{t-1}^{-b_{t}},%
\theta \right) \right. \\
&&\left. K_{t}\left( \widetilde{x}_{t}^{1},dx_{t}^{b_{t}}|\overline{\mathbf{x%
}}_{1:t-1},\mathbf{x}_{t}^{-b_{t}},\widetilde{x}_{t+1}^{Ct+1},\overline{%
\mathbf{a}}_{1:t-2},\mathbf{a}_{t-1}^{-b_{t}},\theta \right) \right\} .
\end{eqnarray*}%
Applying the detailed balance condition in Assumption \ref%
{assumption:detailedbalance} repeatedly gives%
\begin{eqnarray*}
&&\Pi ^{\ast }\left( d\mathbf{u}_{1:T}|\theta \right) = \\
&&\pi \left( dx_{1}^{b_{1}}|\widetilde{x}_{2}^{C_{2}},\theta \right)
\dprod_{t=2}^{T-1}\hat{p}\left( dx_{t}^{b_{t}}|\overline{\mathbf{x}}_{1:t-1},%
\tilde{x}_{t+1}^{C_{t+1}},\overline{\mathbf{a}}_{1:t-2},\theta \right) \pi
\left( dx_{T}^{b_{T}}|\theta \right) \left( \frac{1}{N_{T}}\right) \\
&&\frac{\Psi \left( d\overline{\mathbf{x}}_{1:T},\overline{\mathbf{a}}%
_{1:T-1}|\theta \right) }{M_{1}\left( dx_{1}^{b_{1}}|y_{1},\theta \right)
\dprod_{t=2}^{T}\left\{ W_{t-1}^{a_{t-1}^{b_{t}}}M_{t}\left(
dx_{t}^{b_{t}}|y_{t},x_{t-1}^{a_{t-1}^{b_{t}}},\theta \right) \right\} } \\
&&\dprod_{t=2}^{T}\frac{w_{t-1}^{a_{t-1}^{b_{t}}}f_{t}\left(
x_{t}^{b_{t}}|x_{t-1}^{a_{t-1}^{b_{t}}},\theta \right) }{%
\dsum_{i=1}^{N_{t-1}}w_{t-1}^{i}f_{t}\left( x_{t}^{b_{t}}|x_{t-1}^{i},\theta
\right) } \\
&&\left( \frac{1}{N_{T-1}}\right) \dprod_{j=2}^{C_{T-1}}K_{T-1}\left( 
\widetilde{x}_{T-1}^{j-1},d\widetilde{x}_{T-1}^{j}|\overline{\mathbf{x}}%
_{1:T-2},\mathbf{x}_{T-1}^{-b_{T-1}},x_{T}^{b_{T}},\overline{\mathbf{a}}%
_{1:T-3},\mathbf{a}_{T-2}^{-b_{T-1}},\theta \right) \\
&&K_{T-1}\left( x_{T-1}^{b_{T-1}},d\widetilde{x}_{T-1}^{1}|\overline{\mathbf{%
x}}_{1:T-2},\mathbf{x}_{T-1}^{-b_{T-1}},x_{T}^{b_{T}},\overline{\mathbf{a}}%
_{1:T-3},\mathbf{a}_{T-2}^{-b_{T-1}},\theta \right) \\
&&\dprod_{t=1}^{T-2}\left\{ \left( \frac{1}{N_{t}}\right)
\dprod_{j=2}^{C_{t}}K_{t}\left( \widetilde{x}_{t}^{j-1},d\widetilde{x}%
_{t}^{j}|\overline{\mathbf{x}}_{1:t-1},\mathbf{x}_{t}^{-b_{t}},\widetilde{x}%
_{t+1}^{Ct+1},\overline{\mathbf{a}}_{1:t-2},\mathbf{a}_{t-1}^{-b_{t}},\theta
\right) \right. \\
&&\left. K_{t}\left( x_{t}^{b_{t}},d\widetilde{x}_{t}^{1}|\overline{\mathbf{x%
}}_{1:t-1},\mathbf{x}_{t}^{-b_{t}},\widetilde{x}_{t+1}^{Ct+1},\overline{%
\mathbf{a}}_{1:t-2},\mathbf{a}_{t-1}^{-b_{t}},\theta \right) \right\} .
\end{eqnarray*}%
which is an equivalent measure to $\Psi \left( d\mathbf{u}_{1:T}|\theta
\right) $ defined by (\ref{dist_1}) and (\ref{dist_4}).

To prove Part (iii), we first note that (\ref{dist_1}), (\ref{dist_4}) and (%
\ref{dist_6}) gives%
\begin{align}
& \frac{\Pi \left( d\overline{\mathbf{x}}_{1:T},\overline{\mathbf{a}}%
_{1:T-1},\mathbf{b}_{1:T},d\widetilde{\mathbf{x}}_{1:T-1}|\theta \right) }{%
\Psi \left( d\overline{\mathbf{x}}_{1:T},\overline{\mathbf{a}}_{1:T-1},%
\mathbf{b}_{1:T},d\widetilde{\mathbf{x}}_{1:T-1}|\theta \right) }
\label{MHRatio_lemma_4} \\
& =\left[ \pi \left( d\widetilde{x}_{1}^{C_{1}},\ldots ,d\widetilde{x}%
_{T-1}^{C_{T-1}},dx_{T}^{b_{T}}|\theta \right) \left( \frac{1}{N_{T}}\right)
\right.   \notag \\
& \frac{\Psi \left( d\overline{\mathbf{x}}_{1:T},\overline{\mathbf{a}}%
_{1:T-1}|\theta \right) }{M_{1}\left( dx_{1}^{b_{1}}|y_{1},\theta \right)
\dprod_{t=2}^{T}\left\{ W_{t-1}^{a_{t-1}^{b_{t}}}M_{t}\left(
dx_{t}^{b_{t}}|y_{t},x_{t-1}^{a_{t-1}^{b_{t}}}\right) \right\} }  \notag \\
& \dprod_{t=2}^{T}\frac{w_{t-1}^{a_{t-1}^{b_{t}}}f_{t}\left(
x_{t}^{b_{t}}|x_{t-1}^{a_{t-1}^{b_{t}}},\theta \right) }{%
\dsum_{i=1}^{N_{t-1}}w_{t-1}^{i}f_{t}\left( x_{t}^{b_{t}}|x_{t-1}^{i},\theta
\right) }  \notag \\
& \left( \frac{1}{N_{T-1}}\right) \dprod_{j=2}^{C_{T-1}}K_{T-1}\left( 
\widetilde{x}_{T-1}^{j},d\widetilde{x}_{T-1}^{j-1}|\overline{\mathbf{x}}%
_{1:T-2},\mathbf{x}_{T-1}^{-b_{T-1}},x_{T}^{b_{T}},\overline{\mathbf{a}}%
_{1:T-3},\mathbf{a}_{T-2}^{-b_{T-1}},\theta \right)   \notag \\
& K_{T-1}\left( \widetilde{x}_{T-1}^{1},dx_{T-1}^{b_{T-1}}|\overline{\mathbf{%
x}}_{1:T-2},\mathbf{x}_{T-1}^{-b_{T-1}},x_{T}^{b_{T}},\overline{\mathbf{a}}%
_{1:T-3},\mathbf{a}_{T-2}^{-b_{T-1}},\theta \right)   \notag \\
& \dprod_{t=1}^{T-2}\left\{ \left( \frac{1}{N_{t}}\right)
\dprod_{j=2}^{C_{t}}K_{t}\left( \widetilde{x}_{t}^{j},d\widetilde{x}%
_{t}^{j-1}|\overline{\mathbf{x}}_{1:t-1},\mathbf{x}_{t}^{-b_{t}},\widetilde{x%
}_{t+1}^{Ct+1},\overline{\mathbf{a}}_{1:t-2},\mathbf{a}_{t-1}^{-b_{t}},%
\theta \right) \right.   \notag \\
& \left. \left. K_{t}\left( \widetilde{x}_{t}^{1},dx_{t}^{b_{t}}|\overline{%
\mathbf{x}}_{1:t-1},\mathbf{x}_{t}^{-b_{t}},\widetilde{x}_{t+1}^{Ct+1},%
\overline{\mathbf{a}}_{1:t-2},\mathbf{a}_{t-1}^{-b_{t}},\theta \right)
\right\} \right] /  \notag \\
& \left[ \Psi \left( d\overline{\mathbf{x}}_{1:T},,\overline{\mathbf{a}}%
_{1:T-1}\right) \right. W_{T}^{b_{T}}\widetilde{W}_{T-1}^{b_{T-1}}  \notag \\
& K_{T-1}\left( x_{T-1}^{b_{T-1}},d\widetilde{x}_{T-1}^{1}|\overline{\mathbf{%
x}}_{1:T-2},\mathbf{x}_{T-1}^{-b_{T-1}},x_{T}^{b_{T}},\overline{\mathbf{a}}%
_{1:T-3},\mathbf{a}_{T-2}^{-b_{T-1}},\theta \right)   \notag \\
& \dprod_{j=2}^{C_{T-1}}K_{T-1}\left( \widetilde{x}_{T-1}^{j-1},d\widetilde{x%
}_{T-1}^{j}|\overline{\mathbf{x}}_{1:T-2},\mathbf{x}%
_{T-1}^{-b_{T-1}},x_{T}^{b_{T}},\overline{\mathbf{a}}_{1:T-1},\theta \right) 
\notag \\
& \dprod_{t=1}^{T-2}\left\{ \widetilde{W}_{t}^{b_{t}}K_{t}\left(
x_{t}^{b_{t}},d\widetilde{x}_{t}^{1}|\overline{\mathbf{x}}_{1:t-1},\mathbf{x}%
_{t}^{-b_{t}},\widetilde{x}_{t+1}^{C_{t+1}},\overline{\mathbf{a}}_{1:t-2},%
\mathbf{a}_{t-1}^{-b_{t}},\theta \right) \right.   \notag \\
& \left. \left. \dprod_{j=2}^{C_{t}}K_{t}\left( \widetilde{x}_{t}^{j-1},d%
\widetilde{x}_{t}^{j}|\overline{\mathbf{x}}_{1:t-1},\mathbf{x}_{t}^{-b_{t}},%
\widetilde{x}_{t+1}^{C_{t+1}},\overline{\mathbf{a}}_{1:t-2},\mathbf{a}%
_{t-1}^{-b_{t}},\theta \right) \right\} \right] .  \notag
\end{align}%
To simplify the expression in (\ref{MHRatio_lemma_4}) we note that from the
detailed balance Assumption \ref{assumption:detailedbalance}%
\begin{equation*}
\frac{K_{1}\left( dx_{1}^{\prime }|\mathbf{x}_{1}^{-b_{1}},x_{1},\widetilde{x%
}_{2}^{C_{2}},\theta \right) }{K_{1}\left( dx_{1}|\mathbf{x}%
_{1}^{-b_{1}},x_{1}^{\prime },\widetilde{x}_{2}^{C_{2}},\theta \right) }=%
\frac{p\left( dx_{1}^{\prime }|y_{1},\widetilde{x}_{2}^{C_{2}},\theta
\right) }{p\left( dx_{1}|y_{1},\widetilde{x}_{2}^{C_{2}},\theta \right) },
\end{equation*}%
so%
\begin{eqnarray}
&&\frac{\dprod_{j=2}^{C_{1}}K_{1}\left( \widetilde{x}_{1}^{j},d\widetilde{x}%
_{1}^{j-1}|\mathbf{x}_{1}^{-b_{1}},\widetilde{x}_{2}^{C_{2}},\theta \right)
K_{1}\left( \widetilde{x}_{1}^{1},dx_{1}^{b_{1}}|\mathbf{x}_{1}^{-b_{1}},%
\widetilde{x}_{2}^{C_{2}},\theta \right) }{K_{1}\left( d\widetilde{x}%
_{1}^{1}|\mathbf{x}_{1}^{-b_{1}},x_{1}^{b_{1}},\widetilde{x}%
_{2}^{C_{2}},\theta \right) \dprod_{j=2}^{C_{1}}K_{1}\left( d\widetilde{x}%
_{1}^{j}|\mathbf{x}_{1}^{-b_{1}},\widetilde{x}_{1}^{j-1},\widetilde{x}%
_{2}^{C_{2}},\theta \right) }  \notag \\
&=&\frac{p\left( dx_{1}^{b_{1}}|\widetilde{x}_{2}^{C_{2}},\theta \right) }{%
p\left( d\widetilde{x}_{1}^{C_{1}}|\widetilde{x}_{2}^{C_{2}},\theta \right) }
\notag \\
&=&\frac{g_{1}\left( y_{1}|x_{1}^{b_{1}},\theta \right) f_{2}\left( 
\widetilde{x}_{2}^{C_{2}}|x_{1}^{b_{1}},\theta \right) f_{1}\left(
x_{1}^{b_{1}}|\theta \right) dx_{1}^{b_{1}}}{g_{1}\left( y_{1}|\widetilde{x}%
_{1}^{C_{1}},\theta \right) f_{2}\left( \widetilde{x}_{2}^{C_{2}}|\widetilde{%
x}_{1}^{C_{1}},\theta \right) f_{1}\left( \widetilde{x}_{1}^{C_{1}}|\theta
\right) d\widetilde{x}_{1}^{C_{1}}}.  \label{MHRatio_lemma_5}
\end{eqnarray}

For $t=2,\ldots ,T-2$%
\begin{eqnarray*}
&&\frac{K_{t}\left( x_{t},dx_{t}^{\prime }|\overline{\mathbf{x}}_{1:t-1},%
\mathbf{x}_{t}^{-b_{t}},\widetilde{x}_{t+1}^{Ct+1},\overline{\mathbf{a}}%
_{1:t-2},\mathbf{a}_{t-1}^{-b_{t}},\theta \right) }{K_{t}\left(
x_{t}^{\prime },dx_{t}|\overline{\mathbf{x}}_{1:t-1},\mathbf{x}_{t}^{-b_{t}},%
\widetilde{x}_{t+1}^{Ct+1},\overline{\mathbf{a}}_{1:t-2},\mathbf{a}%
_{t-1}^{-b_{t}},\theta \right) } \\
&=&\frac{\widehat{p}\left( dx_{t}^{\prime }|\overline{\mathbf{x}}_{1:t-1},%
\tilde{x}_{t+1}^{C_{t+1}},\overline{\mathbf{a}}_{1:t-2},\theta \right) }{%
\widehat{p}\left( dx_{t}|\overline{\mathbf{x}}_{1:t-1},\tilde{x}%
_{t+1}^{C_{t+1}},\overline{\mathbf{a}}_{1:t-2},\theta \right) },
\end{eqnarray*}%
so using( \ref{MCMCGenStates_1a})%
\begin{eqnarray}
&&\left\{ \dprod_{j=2}^{C_{t}}K_{t}\left( \widetilde{x}_{t}^{j},d\widetilde{x%
}_{t}^{j-1}|\overline{\mathbf{x}}_{1:t-1},\mathbf{x}_{t}^{-b_{t}},\widetilde{%
x}_{t+1}^{Ct+1},\overline{\mathbf{a}}_{1:t-2},\mathbf{a}_{t-1}^{-b_{t}},%
\theta \right) \right.  \notag \\
&&\left. K_{t}\left( \widetilde{x}_{t}^{1},dx_{t}^{b_{t}}|\overline{\mathbf{x%
}}_{1:t-1},\mathbf{x}_{t}^{-b_{t}},\widetilde{x}_{t+1}^{Ct+1},\overline{%
\mathbf{a}}_{1:t-2},\mathbf{a}_{t-1}^{-b_{t}},\theta \right) \right\} / 
\notag \\
&&\left\{ K_{t}\left( x_{t}^{b_{t}},d\widetilde{x}_{t}^{1}|\overline{\mathbf{%
x}}_{1:t-1},\mathbf{x}_{t}^{-b_{t}},\widetilde{x}_{t+1}^{Ct+1},\overline{%
\mathbf{a}}_{1:t-2},\mathbf{a}_{t-1}^{-b_{t}},\theta \right) \right.  \notag
\\
&&\left. \dprod_{j=2}^{C_{t}}K_{t}\left( \widetilde{x}_{t}^{j-1},d\widetilde{%
x}_{t}^{j}|\overline{\mathbf{x}}_{1:t-1},\mathbf{x}_{t}^{-b_{t}},\widetilde{x%
}_{t+1}^{Ct+1},\overline{\mathbf{a}}_{1:t-2},\mathbf{a}_{t-1}^{-b_{t}},%
\theta \right) \right\}  \notag \\
&=&\frac{\widehat{p}\left( dx_{t}^{b_{t}}|\overline{\mathbf{x}}_{1:t-1},%
\tilde{x}_{t+1}^{C_{t+1}},\overline{\mathbf{a}}_{1:t-2},\theta \right) }{%
\widehat{p}\left( d\widetilde{x}_{t}^{C_{t}}|\overline{\mathbf{x}}_{1:t-1},%
\tilde{x}_{t+1}^{C_{t+1}},\overline{\mathbf{a}}_{1:t-2},\theta \right) } 
\notag \\
&=&\frac{g_{t}\left( y_{t}|x_{t}^{b_{t}},\theta \right) f_{t+1}\left( 
\widetilde{x}_{t+1}^{C_{t+1}}|x_{t}^{b_{t}},\theta \right)
\dsum_{i=1}^{N_{t-1}}W_{t-1}^{i}f_{t}\left( x_{t}^{b_{t}}|x_{t-1}^{i},\theta
\right) dx_{t}^{b_{t}}}{g_{t}\left( y_{t}|\widetilde{x}_{t}^{C_{t}},\theta
\right) f_{t+1}\left( \widetilde{x}_{t+1}^{C_{t+1}}|\widetilde{x}%
_{t}^{C_{t}},\theta \right) \dsum_{i=1}^{N_{t-1}}W_{t-1}^{i}f_{t}\left( 
\widetilde{x}_{t}^{C_{t}}|x_{t-1}^{i},\theta \right) d\widetilde{x}%
_{t}^{C_{t}}}  \notag \\
&=&\frac{g_{t}\left( y_{t}|x_{t}^{b_{t}},\theta \right) f_{t+1}\left( 
\widetilde{x}_{t+1}^{C_{t+1}}|x_{t}^{b_{t}},\theta \right)
\dsum_{i=1}^{N_{t-1}}w_{t-1}^{i}f_{t}\left( x_{t}^{b_{t}}|x_{t-1}^{i},\theta
\right) dx_{t}^{b_{t}}}{g_{t}\left( y_{t}|\widetilde{x}_{t}^{C_{t}},\theta
\right) f_{t+1}\left( \widetilde{x}_{t+1}^{C_{t+1}}|\widetilde{x}%
_{t}^{C_{t}},\theta \right) \dsum_{i=1}^{N_{t-1}}w_{t-1}^{i}f_{t}\left( 
\widetilde{x}_{t}^{C_{t}}|x_{t-1}^{i},\theta \right) d\widetilde{x}%
_{t}^{C_{t}}}.  \label{MHRatio_lemma_6}
\end{eqnarray}%
For $t=T-1$%
\begin{eqnarray*}
&&\frac{K_{T-1}\left( x_{T-1},dx_{T-1}^{\prime }|\overline{\mathbf{x}}%
_{1:T-2},\mathbf{x}_{T-1}^{-b_{T-1}},x_{T}^{b_{T}},\overline{\mathbf{a}}%
_{1},\ldots ,\overline{\mathbf{a}}_{T-3},\mathbf{a}_{T-2}^{-b_{T-1}},\theta
\right) }{K_{T-1}\left( x_{T-1}^{\prime },dx_{T-1}|\overline{\mathbf{x}}%
_{1:T-2},\mathbf{x}_{T-1}^{-b_{T-1}},x_{T}^{b_{T}},\overline{\mathbf{a}}%
_{1},\ldots ,\overline{\mathbf{a}}_{T-3},\mathbf{a}_{T-2}^{-b_{T-1}},\theta
\right) } \\
&=&\frac{\widehat{p}\left( dx_{T-1}^{\prime }|\overline{\mathbf{x}}_{1:T-2},%
\mathbf{x}_{T-1}^{-b_{T-1}},x_{T}^{b_{T}},\overline{\mathbf{a}}_{1},\ldots ,%
\overline{\mathbf{a}}_{T-3},\mathbf{a}_{T-2}^{-b_{T-1}},\theta \right) }{%
\widehat{p}\left( dx_{T-1}|\overline{\mathbf{x}}_{1:T-2},\mathbf{x}%
_{T-1}^{-b_{T-1}},x_{T}^{b_{T}},\overline{\mathbf{a}}_{1},\ldots ,\overline{%
\mathbf{a}}_{T-3},\mathbf{a}_{T-2}^{-b_{T-1}},\theta \right) }.
\end{eqnarray*}%
so using (\ref{MCMCGenStates_1aa})%
\begin{eqnarray}
&&\left\{ \dprod_{j=2}^{C_{T-1}}K_{T-1}\left( \widetilde{x}_{T-1}^{j},d%
\widetilde{x}_{T-1}^{j-1}|\overline{\mathbf{x}}_{1:T-2},\mathbf{x}%
_{T-1}^{-b_{T-1}},x_{T}^{b_{T}},\overline{\mathbf{a}}_{1:T-3},\mathbf{a}%
_{T-2}^{-b_{T-1}},\theta \right) \right.  \notag \\
&&\left. K_{T-1}\left( \widetilde{x}_{T-1}^{1},dx_{T-1}^{b_{T-1}}|\overline{%
\mathbf{x}}_{1:T-2},\mathbf{x}_{T-1}^{-b_{T-1}},x_{T}^{b_{T}},\overline{%
\mathbf{a}}_{1:T-3},\mathbf{a}_{T-2}^{-b_{T-1}},\theta \right) \right\} / 
\notag \\
&&\left\{ K_{T-1}\left( x_{T-1}^{b_{T-1}},d\widetilde{x}_{T-1}^{1}|\overline{%
\mathbf{x}}_{1:T-2},\mathbf{x}_{T-1}^{-b_{T-1}},x_{T}^{b_{T}},\overline{%
\mathbf{a}}_{1:T-3},\mathbf{a}_{T-2}^{-b_{T-1}},\theta \right) \right. 
\notag \\
&&\left. \dprod_{j=2}^{C_{T-1}}K_{T-1}\left( \widetilde{x}_{T-1}^{j-1},d%
\widetilde{x}_{T-1}^{j}|\overline{\mathbf{x}}_{1:T-2},\mathbf{x}%
_{T-1}^{-b_{T-1}},x_{T}^{b_{T}},\overline{\mathbf{a}}_{1:T-3},\mathbf{a}%
_{T-2}^{-b_{T-1}},\theta \right) \right\}  \notag \\
&=&\frac{\widehat{p}\left( dx_{T-1}^{b_{T-1}}|\overline{\mathbf{x}}%
_{1:T-2},x_{T}^{b_{T}},\overline{\mathbf{a}}_{1:T-3},\theta \right) }{%
\widehat{p}\left( d\widetilde{x}_{T-1}^{C_{T-1}}|\overline{\mathbf{x}}%
_{1:T-2},x_{T}^{b_{T}},\overline{\mathbf{a}}_{1:T-3},\theta \right) }
\label{MHRatio_lemma_7} \\
&=&\frac{g_{T-1}\left( y_{T-1}|x_{T-1}^{b_{T-1}},\theta \right) f_{T}\left(
x_{T}^{b_{T}}|x_{T-1}^{b_{T-1}},\theta \right)
\dsum_{i=1}^{N_{T-2}}W_{T-2}^{i}f_{T-1}\left(
x_{T-1}^{b_{T-1}}|x_{T-2}^{i},\theta \right) dx_{T-1}^{b_{T-1}}}{%
g_{T-1}\left( y_{T-1}|\widetilde{x}_{T-1}^{C_{T-1}},\theta \right)
f_{T}\left( x_{T}^{b_{T}}|\widetilde{x}_{T-1}^{C_{T-1}},\theta \right)
\dsum_{i=1}^{N_{T-2}}W_{T-2}^{i}f_{T-1}\left( \widetilde{x}%
_{T-1}^{C_{T-1}}|x_{T-2}^{i},\theta \right) d\widetilde{x}_{T-1}^{C_{T-1}}} 
\notag \\
&=&\frac{g_{T-1}\left( y_{T-1}|x_{T-1}^{b_{T-1}},\theta \right) f_{T}\left(
x_{T}^{b_{T}}|x_{T-1}^{b_{T-1}},\theta \right)
\dsum_{i=1}^{N_{T-2}}w_{T-2}^{i}f_{T-1}\left(
x_{T-1}^{b_{T-1}}|x_{T-2}^{i},\theta \right) dx_{T-1}^{b_{T-1}}}{%
g_{T-1}\left( y_{T-1}|\widetilde{x}_{T-1}^{C_{T-1}},\theta \right)
f_{T}\left( x_{T}^{b_{T}}|\widetilde{x}_{T-1}^{C_{T-1}},\theta \right)
\dsum_{i=1}^{N_{T-2}}w_{T-2}^{i}f_{T-1}\left( \widetilde{x}%
_{T-1}^{C_{T-1}}|x_{T-2}^{i},\theta \right) d\widetilde{x}_{T-1}^{C_{T-1}}}.
\notag
\end{eqnarray}%
Substituting (\ref{MHRatio_lemma_5}), (\ref{MHRatio_lemma_6}) and (\ref%
{MHRatio_lemma_7}) into (\ref{MHRatio_lemma_4}), expanding the terms
involving the normalized weights and rearranging and cancelling the term $%
\Psi \left( d\overline{\mathbf{x}}_{1:T},,\overline{\mathbf{a}}%
_{1:T-1}|\theta \right) $ gives%
\begin{align}
& \frac{\Pi \left( d\overline{\mathbf{x}}_{1:T},,\overline{\mathbf{a}}%
_{1:T-1},\mathbf{b}_{1:T},d\widetilde{\mathbf{x}}_{1:T-1}|\theta \right) }{%
\Psi \left( d\overline{\mathbf{x}}_{1:T},,\overline{\mathbf{a}}_{1:T-1},%
\mathbf{b}_{1:T},d\widetilde{\mathbf{x}}_{1:T-1}|\theta \right) }
\label{MHRatio_lemma_8} \\
& =\frac{\pi \left( \widetilde{x}_{1}^{C_{1}},\ldots ,\widetilde{x}%
_{T-1}^{C_{T-1}},x_{T}^{b_{T}}|\theta \right) \dprod_{t=1}^{T}\left\{ \left(
\dsum_{i=1}^{N_{t}}w_{t}^{i}\right) \left( \frac{1}{N_{t}}\right) \right\} }{%
M_{1}\left( x_{1}^{b_{1}}|y_{1},\theta \right) \dprod_{t=2}^{T}\left\{
\left( w_{t-1}^{a_{t-1}^{b_{t}}}\right) M_{t}\left(
x_{t}^{b_{t}}|y_{t},x_{t-1}^{a_{t-1}^{b_{t}}},\theta \right) \right\} }%
\dprod_{t=2}^{T}\frac{w_{t-1}^{a_{t-1}^{b_{t}}}f_{t}\left(
x_{t}^{b_{t}}|x_{t-1}^{a_{t-1}^{b_{t}}},\theta \right) }{%
\dsum_{i=1}^{N_{t-1}}w_{t-1}^{i}f_{t}\left( x_{t}^{b_{t}}|x_{t-1}^{i},\theta
\right) }  \notag \\
& \left[ \left( w_{T}^{b_{T}}\right) \left( \frac{w_{T-1}^{b_{T-1}}f_{T}%
\left( x_{T}^{b_{T}}|x_{T-1}^{b_{T-1}},\theta \right) }{%
\dsum_{i=1}^{N_{T-1}}w_{T-1}^{i}f_{T}\left( x_{T}^{b_{T}}|x_{T-1}^{i},\theta
\right) }\right) \dprod_{t=1}^{T-2}\frac{w_{t}^{b_{t}}f_{t+1}\left( 
\widetilde{x}_{t+1}^{C_{t+1}}|x_{t}^{b_{t}},\theta \right) }{%
\dsum_{i=1}^{N_{t}}w_{t}^{i}f_{t+1}\left( \widetilde{x}%
_{t+1}^{C_{t+1}}|x_{t}^{i},\theta \right) }\right] ^{-1}  \notag \\
& \frac{g_{1}\left( y_{1}|x_{1}^{b_{1}},\theta \right) f_{2}\left( 
\widetilde{x}_{2}^{C_{2}}|x_{1}^{b_{1}},\theta \right) f_{1}\left(
x_{1}^{b_{1}}|\theta \right) }{g_{1}\left( y_{1}|\widetilde{x}%
_{1}^{C_{1}},\theta \right) f_{2}\left( \widetilde{x}_{2}^{C_{2}}|\widetilde{%
x}_{1}^{C_{1}},\theta \right) f_{1}\left( \widetilde{x}_{1}^{C_{1}}|\theta
\right) }  \notag \\
& \dprod_{t=2}^{T-2}\left\{ \frac{g_{t}\left( y_{t}|x_{t}^{b_{t}},\theta
\right) f_{t+1}\left( \widetilde{x}_{t+1}^{C_{t+1}}|x_{t}^{b_{t}},\theta
\right) \dsum_{i=1}^{N_{t-1}}w_{t-1}^{i}f_{t}\left(
x_{t}^{b_{t}}|x_{t-1}^{i},\theta \right) }{g_{t}\left( y_{t}|\widetilde{x}%
_{t}^{C_{t}},\theta \right) f_{t+1}\left( \widetilde{x}_{t+1}^{C_{t+1}}|%
\widetilde{x}_{t}^{C_{t}},\theta \right)
\dsum_{i=1}^{N_{t-1}}w_{t-1}^{i}f_{t}\left( \widetilde{x}%
_{t}^{C_{t}}|x_{t-1}^{i},\theta \right) }\right\}  \notag \\
& \frac{g_{T-1}\left( y_{T-1}|x_{T-1}^{b_{T-1}},\theta \right) f_{T}\left(
x_{T}^{b_{T}}|x_{T-1}^{b_{T-1}},\theta \right)
\dsum_{i=1}^{N_{T-2}}w_{T-2}^{i}f_{T-1}\left(
x_{T-1}^{b_{T-1}}|x_{T-2}^{i},\theta \right) }{g_{T-1}\left( y_{T-1}|%
\widetilde{x}_{T-1}^{C_{T-1}},\theta \right) f_{T}\left( x_{T}^{b_{T}}|%
\widetilde{x}_{T-1}^{C_{T-1}},\theta \right)
\dsum_{i=1}^{N_{T-2}}w_{T-2}^{i}f_{T-1}\left( \widetilde{x}%
_{T-1}^{C_{T-1}}|x_{T-2}^{i},\theta \right) }  \notag \\
& =\frac{\pi \left( \widetilde{x}_{1}^{C_{1}},\ldots ,\widetilde{x}%
_{T-1}^{C_{T-1}},x_{T}^{b_{T}}|\theta \right) \dprod_{t=1}^{T}\left\{ \left(
\dsum_{i=1}^{N_{t}}w_{t}^{i}\right) \left( \frac{1}{N_{t}}\right) \right\} }{%
M_{1}\left( x_{1}^{b_{1}}|y_{1},\theta \right) \dprod_{t=2}^{T}M_{t}\left(
x_{t}^{b_{t}}|y_{t},x_{t-1}^{a_{t-1}^{b_{t}}},\theta \right) }%
\dprod_{t=2}^{T}f_{t}\left( x_{t}^{b_{t}}|x_{t-1}^{a_{t-1}^{b_{t}}},\theta
\right)  \notag \\
& \left( \dprod_{t=1}^{T}w_{t}^{b_{t}}\right) ^{-1}\frac{g_{1}\left(
y_{1}|x_{1}^{b_{1}},\theta \right) f_{1}\left( x_{1}^{b_{1}}|\theta \right) 
}{g_{1}\left( y_{1}|\widetilde{x}_{1}^{C_{1}},\theta \right) f_{2}\left( 
\widetilde{x}_{2}^{C_{2}}|\widetilde{x}_{1}^{C_{1}},\theta \right)
f_{1}\left( \widetilde{x}_{1}^{C_{1}}|\theta \right) }  \notag \\
& \left\{ \dprod_{t=2}^{T-2}\frac{g_{t}\left( y_{t}|x_{t}^{b_{t}},\theta
\right) }{g_{t}\left( y_{t}|\widetilde{x}_{t}^{C_{t}},\theta \right)
f_{t+1}\left( \widetilde{x}_{t+1}^{C_{t+1}}|\widetilde{x}_{t}^{C_{t}},\theta
\right) }\right\} \frac{g_{T-1}\left( y_{T-1}|x_{T-1}^{b_{T-1}},\theta
\right) }{g_{T-1}\left( y_{T-1}|\widetilde{x}_{T-1}^{C_{T-1}},\theta \right)
f_{T}\left( x_{T}^{b_{T}}|\widetilde{x}_{T-1}^{C_{T-1}},\theta \right) } 
\notag
\end{align}%
Equation (\ref{smc_2}) implies that%
\begin{equation}
w_{1}^{b_{1}}M_{1}\left( x_{1}^{b_{1}}|y_{1},\theta \right) =g_{1}\left(
y_{1}|x_{1}^{b_{1}},\theta \right) f_{1}\left( x_{1}^{b_{1}}|\theta \right) ,
\label{MHRatio_lemma_9}
\end{equation}%
and (\ref{smc_3}) implies that for $t=2,\ldots ,T$%
\begin{equation}
w_{t}^{b_{t}}M_{t}\left(
x_{t}^{b_{t}}|y_{t},x_{t-1}^{a_{t-1}^{b_{t}}},\theta \right) =g_{t}\left(
y_{t}|x_{t}^{b_{t}},\theta \right) f_{t}\left(
x_{t}^{b_{t}}|x_{t-1}^{a_{t-1}^{b_{t}}},\theta \right) .
\label{MHRatio_lemma_10}
\end{equation}%
Substituting (\ref{MHRatio_lemma_9}) and (\ref{MHRatio_lemma_10})\ into (\ref%
{MHRatio_lemma_8}) gives%
\begin{align*}
& \frac{\Pi \left( d\overline{\mathbf{x}}_{1:T},,\overline{\mathbf{a}}%
_{1:T-1},\mathbf{b}_{1:T},d\widetilde{\mathbf{x}}_{1:T-1}|\theta \right) }{%
\Psi \left( \overline{d\mathbf{x}}_{1:T},,\overline{\mathbf{a}}_{1:T-1},%
\mathbf{b}_{1:T},d\widetilde{\mathbf{x}}_{1:T-1}|\theta \right) } \\
& =\frac{\pi \left( \widetilde{x}_{1}^{C_{1}},\ldots ,\widetilde{x}%
_{T-1}^{C_{T-1}},x_{T}^{b_{T}}|\theta \right) \dprod_{t=1}^{T}\left\{ \left(
\dsum_{i=1}^{N_{t}}w_{t}^{i}\right) \left( \frac{1}{N_{t}}\right) \right\} }{%
\left\{ \dprod_{t=1}^{T-1}g_{t}\left( y_{t}|\widetilde{x}_{t}^{C_{t}},\theta
\right) \right\} g_{T}\left( y_{T}|x_{T}^{bT},\theta \right) f_{1}\left( 
\widetilde{x}_{1}^{C_{1}}|\theta \right) \dprod_{t=1}^{T-2}f_{t+1}\left( 
\widetilde{x}_{t+1}^{C_{t+1}}|\widetilde{x}_{t}^{C_{t}},\theta \right)
f_{T}\left( x_{T}^{b_{T}}|\widetilde{x}_{T-1}^{C_{T-1}},\theta \right) } \\
& =\frac{\dprod_{t=1}^{T}\left\{ \left( \dsum_{i=1}^{N_{t}}w_{t}^{i}\right)
\left( \frac{1}{N_{t}}\right) \right\} }{p\left( y_{1:T}|\theta \right) },
\end{align*}%
as required.

\textit{Proof of Lemma \ref{PG_lemma_2}.}

Let%
\begin{equation*}
h^{\ast }\left( \overline{\mathbf{x}}_{1:T},\overline{\mathbf{a}}%
_{1:T-1}|\theta \right) =\frac{\dprod_{t=1}^{T}\left\{ \left(
\dsum_{i=1}^{N_{t}}w_{t-1}^{i}\right) \left( \frac{1}{N_{t}}\right) \right\} 
}{p\left( \mathbf{y}_{1:T}|\theta \right) }.
\end{equation*}%
From (\ref{dist_8}),

\begin{equation}
\Pi \left( d\overline{\mathbf{x}}_{1:T},\overline{\mathbf{a}}_{1:T-1},%
\mathbf{b}_{1:T},d\widetilde{\mathbf{x}}_{1:T-1}|\theta \right) =h^{\ast
}\left( \overline{\mathbf{x}}_{1:T},\overline{\mathbf{a}}_{1:T-1}|\theta
\right) \Psi \left( d\overline{\mathbf{x}}_{1:T},\overline{\mathbf{a}}%
_{1:T-1},\mathbf{b}_{1:T},d\widetilde{\mathbf{x}}_{1:T-1}|\theta \right) .
\label{PG_lemma_2_1}
\end{equation}%
Integrating (\ref{PG_lemma_2_1}) over $\mathbf{b}_{1:T},\widetilde{\mathbf{x}%
}_{1:T-1}$ shows that the marginal distributions of $\Pi \left( d\overline{%
\mathbf{x}}_{1:T},\overline{\mathbf{a}}_{1:T-1}\right) $ and $\Psi \left( d%
\overline{\mathbf{x}}_{1:T},\overline{\mathbf{a}}_{1:T-1}\right) $ satisfy%
\begin{equation*}
\Pi \left( d\overline{\mathbf{x}}_{1:T},\overline{\mathbf{a}}_{1:T-1}|\theta
\right) =h^{\ast }\left( \overline{\mathbf{x}}_{1:T},\overline{\mathbf{a}}%
_{1:T-1}|\theta \right) \Psi \left( d\overline{\mathbf{x}}_{1:T},\overline{%
\mathbf{a}}_{1:T-1}|\theta \right) .
\end{equation*}%
Hence the conditional distribution of $\Pi \left( \mathbf{b}_{1:T},d%
\widetilde{\mathbf{x}}_{1:T-1}|\overline{\mathbf{x}}_{1:T},\overline{\mathbf{%
a}}_{1:T-1},\theta \right) $is given by%
\begin{eqnarray*}
\Pi \left( \mathbf{b}_{1:T},d\widetilde{\mathbf{x}}_{1:T-1}|\overline{%
\mathbf{x}}_{1:T},\overline{\mathbf{a}}_{1:T-1},\theta \right) &=&\frac{\Pi
\left( d\overline{\mathbf{x}}_{1:T},\overline{\mathbf{a}}_{1:T-1},\mathbf{b}%
_{1:T},d\widetilde{\mathbf{x}}_{1:T-1}|\theta \right) }{\Pi \left( d%
\overline{\mathbf{x}}_{1:T},\overline{\mathbf{a}}_{1:T-1}|\theta \right) } \\
&=&\frac{\Psi \left( d\overline{\mathbf{x}}_{1:T},\overline{\mathbf{a}}%
_{1:T-1},\mathbf{b}_{1:T},d\widetilde{\mathbf{x}}_{1:T-1}|\theta \right) }{%
\Psi \left( d\overline{\mathbf{x}}_{1:T},\overline{\mathbf{a}}%
_{1:T-1}|\theta \right) },
\end{eqnarray*}%
which shows that%
\begin{equation*}
\Pi \left( \mathbf{b}_{1:T},d\widetilde{\mathbf{x}}_{1:T-1}|\overline{%
\mathbf{x}}_{1:T},\overline{\mathbf{a}}_{1:T-1},\theta \right) =\Psi \left( 
\mathbf{b}_{1:T},d\widetilde{\mathbf{x}}_{1:T-1}|\overline{\mathbf{x}}_{1:T},%
\overline{\mathbf{a}}_{1:T-1},\theta \right) ,
\end{equation*}%
as required.

\textit{Proof of Lemma \ref{PGAlgorithm_lemma_1}.}

The proof is similar to the proof of Part (i) of Lemma \ref{MHRatio_lemma}.
Note, however, that the order is reversed since the algorithm starts by
generating from the simplest marginal distributions and then adds variables
by generating from their conditional distributions.

Equation (\ref{MHRatio_lemma_0c}) in the proof of Lemma \ref{MHRatio_lemma}
derives the expressions in Step 1 for the case when $t=1$. Similarly,
equations (\ref{MHRatio_lemma_0a1}) and (\ref{MHRatio_lemma_0b}) derive the
expressions in Step 1 for the cases when $t=2,\ldots ,T-1$. Finally,
equation (\ref{MHRatio_lemma_0a}) derives the expressions in Steps 2 and 3.

\textit{Proof of Theorem }\ref{PGsmoothing_theorem_1}.

Sampling Scheme \ref{PGsmoothing_sscheme} is a Gibbs sampler targeting $\Pi
\left( d\mathbf{U}_{1:T}|\theta \right) $ by construction, so it is
sufficient to show irreducibility and aperiodicity of the Markov chain.

From Step 2, the marginal process involving $\overline{\mathbf{X}}_{1:T},%
\overline{\mathbf{A}}_{1:T-1},\mathbf{B}_{1:T}$ is a Markov chain. From
Assumption \ref{assumption:productkernel}, the accessible sets of this
marginal chain are the same as the assessible sets of the Particle Gibbs
sampler of \cite{lindstenschon2012} with fixed parameters $\theta $. From
Assumption \ref{assumption:support}, Theorem 1 of \cite{lindstenschon2012}
applies with fixed parameters $\theta $, and hence the marginal chain
involving $\overline{\mathbf{X}}_{1:T},\overline{\mathbf{A}}_{1:T-1},\mathbf{%
B}_{1:T}$ is irreducible and aperiodic.

From Step 2 of Sampling Scheme \ref{PGsmoothing_sscheme}, $\widetilde{%
\mathbf{X}}_{1:T-1}$ is generated from $\Pi \left( d\widetilde{\mathbf{X}}%
_{1:T-1}|\overline{\mathbf{X}}_{1:T},\overline{\mathbf{A}}_{1:T-1},\mathbf{B}%
_{1:T},\theta \right) $ and hence the full chain involving $\overline{%
\mathbf{X}}_{1:T},\overline{\mathbf{A}}_{1:T-1},\mathbf{B}_{1:T},\widetilde{%
\mathbf{X}}_{1:T-1}$ is also irreducible and aperiodic.

\textit{Proof of Theorem }\ref{Hybrid_sscheme_theorem_1}.

We first note that if there are PMMH steps in Part 1 of Sampling Scheme \ref%
{Hybrid_sscheme} then irreducibility and aperiodiciy follows from Lemma \ref%
{MHRatio_lemma} (ii).

Suppose there are no PMMH steps in Part 1 of Sampling Scheme \ref%
{Hybrid_sscheme} and the resulting Markov chain is reducible or periodic.
This implies that for any fixed value of $\theta \in \Theta $ the Markov
chain for the particle Gibbs smoother in Sampling Scheme \ref%
{PGsmoothing_sscheme} \ is also reducible or periodic, contradicting Theorem %
\ref{PGsmoothing_theorem_1}.

\end{document}